# Ultrafast non-destructive measurement of the quantum state of light using free electrons

Alexey Gorlach[1], Aviv Karnieli[2], Raphael Dahan[1], Eliahu Cohen[3], Avi Pe'er[3], and Ido Kaminer[1]

1. Solid State Institute, Technion-Israel Institute of Technology, Haifa 32000, Israel
2. Sackler School of Physics, Faculty of Exact Sciences, Tel Aviv University, Tel Aviv, 69978 Israel.
3. Department of Physics and BINA Center of Nanotechnology, Bar-Ilan University, 52900 Ramat-Gan, Israel

kaminer@technion.ac.il

**Since the birth of quantum optics, the measurement of quantum states of nonclassical light has been of tremendous importance for advancement in the field. To date, conventional detectors such as photomultipliers, avalanche photodiodes, and superconducting nanowires, all rely at their core on linear excitation of *bound electrons* with light, posing fundamental restrictions on the detection. In contrast, the interaction of *free electrons* with light in the context of quantum optics is highly nonlinear and offers exciting possibilities. The first experiments that promoted this direction appeared over the past decade as part of photon-induced nearfield electron microscopy (PINEM), wherein free electrons are capable of high-order multi-photon absorption and emission. Here we propose using *free electrons* for quantum-optical detection of the complete quantum state of light. We show how the precise control of the electron before and after its interaction with quantum light enables to extract the photon statistics and implement full quantum state tomography using PINEM. This technique can reach sub-attosecond time resolutions, measure temporal coherence of any degree (e.g., $g^{(1)}$, $g^{(2)}$), and simultaneously detect large numbers of photons with each electron. Importantly, the interaction of the electron with light is *non-destructive*, thereby leaving the photonic state (modified by the interaction) intact, which is conceptually different from conventional detectors. By using a pulse of multiple electrons, we envision how PINEM quantum detectors could achieve a *single-shot* measurement of the complete state of quantum light, even for non-reproducible emission events. Altogether, our work paves the way to novel kinds of photodetectors that utilize the ultrafast duration, high nonlinearity, and non-destructive nature of electron–light interactions.**

**Section I - Introduction**

Quantum optics [1,2] has seen significant advancements in the past decades, allowing for fundamental tests of quantum mechanics [3,4], observations of nonclassical states of light [5–7], and realization of quantum technologies [8–13]. Much of these achievements were made possible thanks to the development of new concepts for measuring photons: with examples ranging from the Hanbury Brown-Twiss (HBT) experiment [14] and coincidence measurements [4], to photon-number resolving detectors [15,16] and homodyne detection [17,18], used for quantum state tomography [2,19–23], which enables the reconstruction of the entire photonic state. All current quantum optical detectors are based on the interaction of light with *bound-electron* systems, such as avalanche photodiodes [24,25], superconducting nanowires [26], and photomultiplier tubes [27]. While many of these platforms utilize electrons bound in solids, we also see electrons bound in atoms (placed in optical cavities) proposed for precise quantum optical detectors [28–32].

The state-of-the-art quantum-optical detectors are based on the linear interaction between bound electrons and light through dipole transitions in the matter. They are therefore inherently governed by the one photon - one electron relation of first-order perturbation theory and the Fermi golden rule. For example, the rich information contained in multi-photon light states often requires the simultaneous usage of multiple detectors [13,33]. Existing quantum-optical detection schemes are also limited in their detection rate and bandwidth due to the response time of electronic components. Great efforts are being invested in overcoming these limitations, for example, using optical nonlinearities [34,35]. Our work considers an entirely new type of interaction for quantum-optical detection, between light and *free electrons*, which we show can be harnessed to improve certain properties of quantum optical detection.

The free-electron–light interaction can be made especially strong using nanophotonic structures and other optical elements coupling the light to the electron. These ideas were promoted by theoretical and experimental advances in the area of photon-induced nearfield electron microscopy (PINEM) [36–44], which showed how free-electron wavepackets could coherently absorb and emit a discrete number of photons at ultrafast time resolution. The free-electron–light interaction strength is maximized when on resonance, i.e. when phase-matching conditions are satisfied. These conditions are at the core of laser-driven particle acceleration [45,46], such as dielectric laser accelerators (DLA) [47,48]. The same phase-matching conditions were recently described and demonstrated [42] in a PINEM experiment, reaching record free-electron–light interaction strengths. These recent demonstrations serve as a strong motivation for our work below.

Here we propose a novel paradigm for quantum-optical detection based on the interaction of light with free electrons. This paradigm is based on the free-electron–light entanglement [49] when the information about the light's state is contained in the electron energy spectrum. We find that free-electron–light interactions enable the full reconstruction of the quantum state of light (e.g., Wigner function of the light [2,50]). Particularly, we show how to directly obtain the photon number statistics from the PINEM interaction, using a new theory and an analytic algorithm presented in this work. Then we develop a more advanced scheme of homodyne detection by free electrons, which gives rise to full quantum state tomography.

We present several intriguing properties of the general concept of free-electron quantum-optical detection (FEQOD): They are sensitive to the temporal optical coherence, have potentially unlimited sensitivity to simultaneous photodetection events, and unprecedented temporal resolution thanks to the ultrashort electron probe duration [51,52], which could reach attosecond and even sub-attosecond levels [40,53–55]. This way, ultrashort electron pulses would enable to increase the bandwidth/temporal resolution by three orders of magnitude compared with conventional number-resolving detectors [9,15,16,56,57] (that are limited by the timescale of electronics). Moreover, unlike all existing platforms for quantum-optical detection, FEQOD performs the detection without destroying the photonic state and leaving the photonic state (modified by the interaction) intact, a feature that can potentially be used for *non-destructive* measurements.

From the point of view of electron microscopy and spectroscopy, our proposal of FEQOD can also be seen as part of the expanding toolbox of analytical electron microscopy [36–44] and spectroscopy [58]. Advances in this field include free-electron-driven bunched [59] and anti-bunched [60] photon sources in time-resolved cathodoluminescence [58,61], as well as recent ideas for using PINEM to create and modify quantum states of light [62]. Our work is part of this bigger thrust of free-electron quantum optics but differs from all previous approaches in a fundamental way: We place free electrons on the opposite side of the quantum optical experiment – not to excite quantum states of light, but as detectors of an *externally generated* quantum light.

**Section II – The quantum theory of free-electron quantum-optical detection (FEQOD)**

The first theoretical models of PINEM [37,38] considered quantized free electrons and classical electromagnetic fields. In these models, all the information about the quantum properties of light is lost. Nevertheless, the theory precisely describes the energy-loss spectrum of electrons in PINEM for strong coherent states (e.g., when illuminating with a laser). Recently, a model of PINEM interaction with the quantized electromagnetic field (QPINEM) has been formulated [49,63]. At the same time, the intensities used in PINEM experiments are gradually decreasing by using more efficient optical structures [41,42,48,64]. These advances give reason to believe that interaction efficiencies will reach the single-photon–single-electron interaction regime. Therefore, an important milestone is to obtain an analytical description of the quantum light–free-electron interaction for any photonic states.

Below we derive the general formalism for the QPINEM interaction and detail its potential use for FEQOD. Our formalism goes beyond previous papers [49,63] by providing analytical expressions for QPINEM of an arbitrary number of photons. We then utilize these results to present the concept of FEQOD for the first time (Figure 1). Here, a free electron interacts with the light pulse through an optical or nanophotonic structure that facilitates electron–photon coupling with a coupling constant $g_q$ (examples of experimental structures are given in Section V). The primary purpose of this structure is to reach the maximal interaction strength, which generally requires confining the light [41,43], and reaching phase-matched interactions with the free electron [42,49].

As we show in the next sections, it is valuable to separate the QPINEM interaction into two regimes: weak ($|g_q| \ll 1$) and strong ($|g_q| \sim 1$) interaction. Each regime can be used for a different application: The strong-interaction regime enables the reconstruction of a weak field (potentially even single-photon sources), and the weak-interaction regime enables us to perform *non-destructive measurements* of stronger fields.

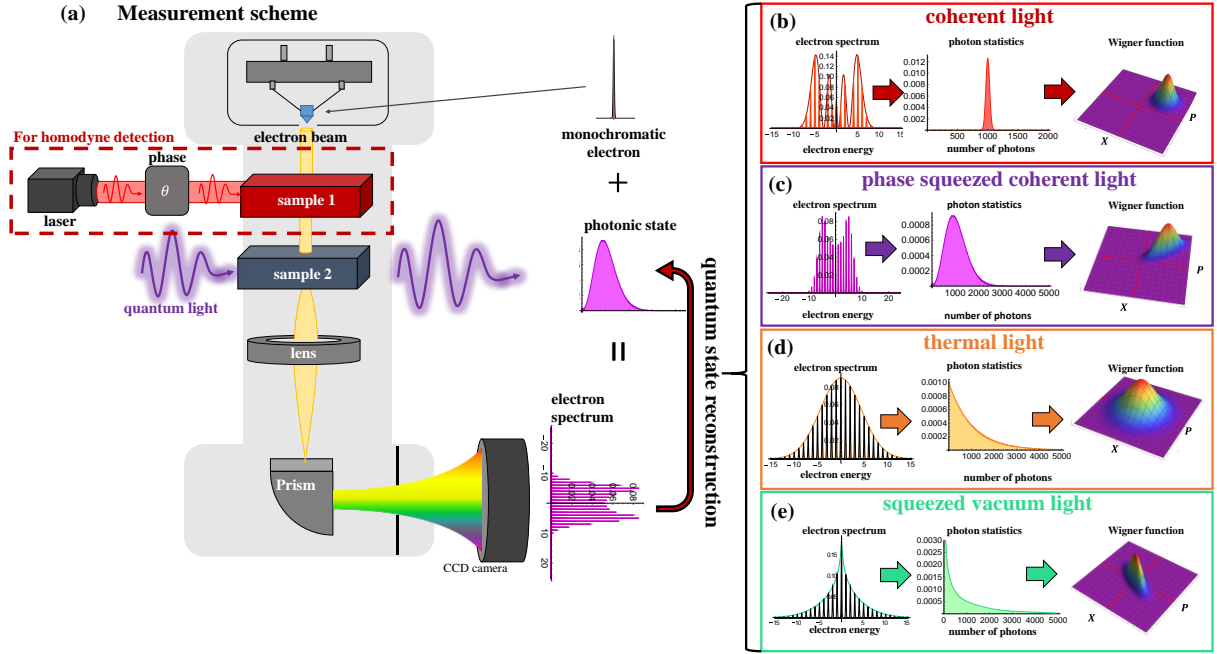

Figure 1. Principle scheme of free-electron quantum-optical detection (FEQOD): reconstructing the quantum states of light from the electron energy spectra for interaction with different types of photonic states. (a) A monoenergetic free-electron pulse interacts with a light pulse through the samples that include an optical/nanophotonic structure that facilitates strong electron–photon coupling. After the interaction, the electrons are measured by electron energy-loss spectroscopy (EELS). We can measure the photon statistics with just one point of interaction (without sample 1). For the reconstruction of the entire photonic state, it is necessary to have an additional point of interaction (sample 1), which can be driven by a laser (marked by the dashed red frame). By changing the interaction parameters (e.g., the relative phase $\theta$) and re-measuring the energy spectra multiple times, it is possible to reconstruct all properties of the quantum state. (b)-(e) Electron energy spectra, corresponding photon number distributions, and Wigner functions for (b) coherent state, (c) phase-squeezed state, (d) thermal state, and (e) squeezed-vacuum state. The spectra in this example were calculated for the same coupling constant $g_q = 0.1$ and for the same average number of photons $\langle n \rangle = 1000$. The same principle works for other parameters (including larger numbers of photons and small $g_q$).

To describe the post-interaction electron spectrum, we adopt the QPINEM theory [49,63]. The interaction between light and a free electron is described by the scattering matrix:

$$S = \exp[g_q b a^\dagger - g_q^* b^\dagger a], \quad (1)$$

where $g_q$ is the complex dimensionless coupling constant; $b, b^\dagger$ are the energy ladder operators for the free-electron [footnote: the $b$ operator represents an electron losing energy $\hbar\omega$ in the paraxial approximation and is defined in the coordinate representation by $b \equiv e^{-i\frac{\omega}{v}z}$, with $v$ denoting the electron velocity and $z$ its longitudinal position; this operator has the following commutation relation $[b, b^\dagger] = 0$]; $a$ and $a^\dagger$ are annihilation and creation operators of the photonic field, respectively. The joint density matrix $|E_0\rangle\langle E_0| \otimes \rho_{\text{ph}}$ describes the initial state of the system, where $|E_0\rangle$ is the state of the initial electron with energy $E_0$, and $\rho_{\text{ph}}$ is the density matrix of the photonic field.

From Eq. (1), we derive the exact analytical formula describing the electron energy spectrum, resulting in:

$$P_k = \begin{cases} |g_q|^{2k} e^{|g_q|^2} \text{Tr}\left[\rho_{\text{ph}} \cdot \frac{\hat{n}!}{(|k|!)^2(\hat{n}-|k|)!} \left|{}_1F_1\left(1+\hat{n}, 1+|k|, -|g_q|^2\right)\right|^2\right], & k > 0 \\ |g_q|^{2|k|} e^{|g_q|^2} \text{Tr}\left[\rho_{\text{ph}} \cdot \frac{(\hat{n}+|k|)!}{\hat{n}!(|k|!)^2} \left|{}_1F_1\left(1+|k|+\hat{n}, 1+|k|, -|g_q|^2\right)\right|^2\right], & k < 0 \end{cases}, \quad (2)$$

where $P_k$ is the probability for the electron to absorb or emit $k$ photons (absorption is given by positive sign and emission is given by negative sign); $\hat{n} = a^\dagger a$ is the photon number operator; ${}_1F_1$ is the hypergeometric function of the operator $\hat{n}$; $\text{Tr}[\rho_{\text{ph}} \ldots]$ is a trace over the photonic state. Eq. (2) is applicable for any value of $g_q$, and is essential to describe the strong interaction between an electron and light. In the regime of small-to-moderate $g_q$ (roughly $|g_q| \lesssim 0.3$), Eq. (2) can be simplified:

$$P_k = \sum_{n=0}^{\infty} p_n \left|J_k\left(2|g_q|\sqrt{n}\right)\right|^2, \quad (3)$$

where $p_n = \text{Tr}[\rho_{\text{ph}}\hat{n}]$ is the probability of finding $n$ photons in the field, and $J_k$ is the Bessel function of order $k$. We note that there is a wide range of parameters for which $g_q$ is both large enough to make the PINEM observable for just several dozens of photons and also small enough so that Eq. (3) is a good approximation (as in the examples in our figures). The derivation of Eqs. (2-*3*) and analysis of their accuracy are detailed in Supplementary Material (SM) Section III. Eqs. (2-3) show that the electron spectrum strongly depends on the photon statistics – which is the key for our reconstruction method in the next sections. Figure 2 presents several examples of the connection between the electron spectrum and photon statistics.

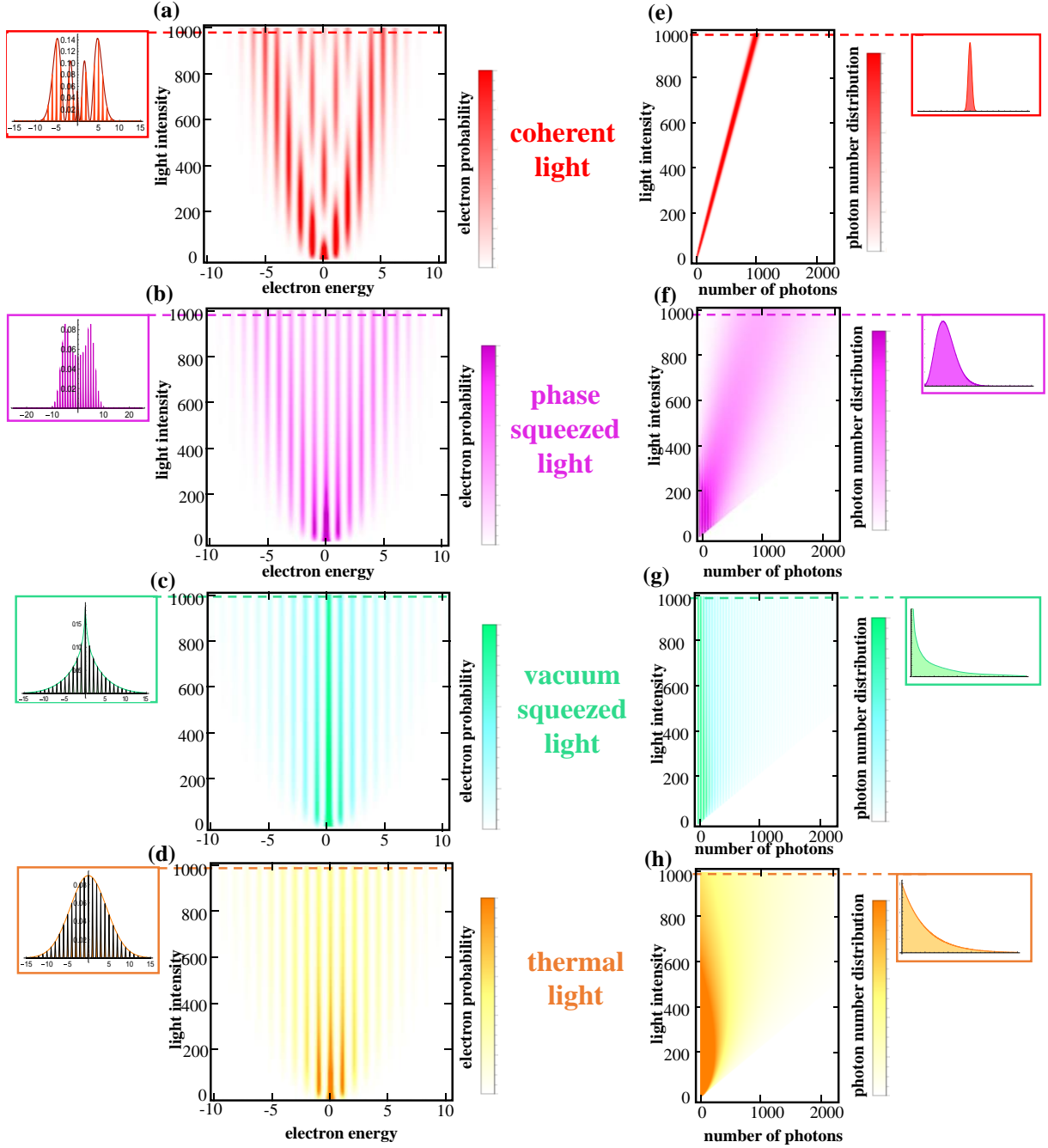

**Figure 2. The correspondence between the electron energy spectra and photon statistics for different photonic states**. **(a-d)** Maps of electron spectra as a function of light intensity (i.e., different average number of photons) for **(a)** coherent, **(b)** phase-squeezed, **(c)** vacuum-squeezed, and **(d)** thermal states. The horizontal axis represents the electron energy spectrum in units of $\hbar\omega$, the vertical axis represents $\langle n \rangle$, the color intensity shows the probability of the electron to obtain specific energies. Insets connected by dashed lines show the spectrum for $\langle n \rangle = 1000$. Color maps **(e-h)** show the photon statistics as a function of $\langle n \rangle$ for **(e)** coherent, **(f)** phase-squeezed, **(g)** vacuum-squeezed, and **(h)** thermal states. The results in this figure match the known analytical expressions for coherent PINEM[37,65], thermal QPINEM[63], and a new analytical expression we found for vacuum-squeezed QPINEM (SM Section IV). All the spectra were calculated for the same coupling constant $g_q = 0.1$.

**Section III – Reconstruction of photon statistics from the electron energy spectrum**

We now derive the fundamental connection between photon statistics and the electron energy spectrum after the interaction. This connection enables to extract the photon statistics and shows what processes can affect the accuracy of the measurement. We consider strong- and weak-interaction regimes between light and free electrons and discuss their potential applications.

The electron energy probabilities $P_k$ (Eq. (2) or approximated Eq. (3)), can be expressed in terms of the photon number operator moments (see SM, Section V):

$$P_k = \sum_{m=|k|}^{\infty} c_{km} \langle n^m \rangle, \tag{4}$$

where $\langle n^m \rangle = \sum_{n=0}^{\infty} p_n n^m$ is the $m^{\text{th}}$ moment of the photon number operator. The coefficients $c_{km}$ have a simple analytical form in the case of Eq. (3):

$$c_{km} = \frac{(-1)^{m-|k|} |g_{\text{q}}|^{2m}}{(m-|k|)!(m+|k|)!} \frac{(2m)!}{(m!)^2};$$

Hence, each peak in the QPINEM spectrum contains information about all the moments of the optical field. Note that higher moments affect the spectrum less because the contribution is proportional to $|g_{\text{q}}|^{2m}$ and the coupling constant $g_{\text{q}}$ is usually smaller than unity. This dependence leads to the conclusion that the absolute value of $g_{\text{q}}$ affects the accuracy of the photon statistics reconstruction, which emphasizes a certain advantage of large $g_{\text{q}}$ values.

Now, we invert Eq. (4) and extract the photon number moments from the measured electron probabilities $P_k$ (SM, Section V):

$$\langle n^m \rangle = \sum_{k=m}^{\infty} d_{mk} P_k, \tag{5}$$

where $d_{mk}$ are the inverted matrix coefficients, which for the case of Eq. (3) have a simple analytical form:

$$d_{mk} = 2^{\lfloor \frac{k-m+1}{2} \rfloor} \cdot \frac{k \cdot m!}{|g_{\text{q}}|^{2m}} \cdot \frac{\left(m-1+\lfloor \frac{k-m+1}{2} \rfloor\right)!\left(2m+2\lfloor \frac{k-m}{2} \rfloor -1\right)!!}{(k-m)!(2m-1)!!},$$

where the square brackets $\lfloor ... \rfloor$ stand for the integer part of the number (i.e., the floor function). Notably, the photon statistics can be reconstructed from the moments; however, the complete reconstruction of the quantum state of light requires not only knowledge of the photon number but also the phase. It can be achieved using an additional point of interaction with a coherent laser (i.e., local oscillator). The reconstruction of the entire quantum state is further discussed in the next section. The general concept is shown in Figure 3(a).

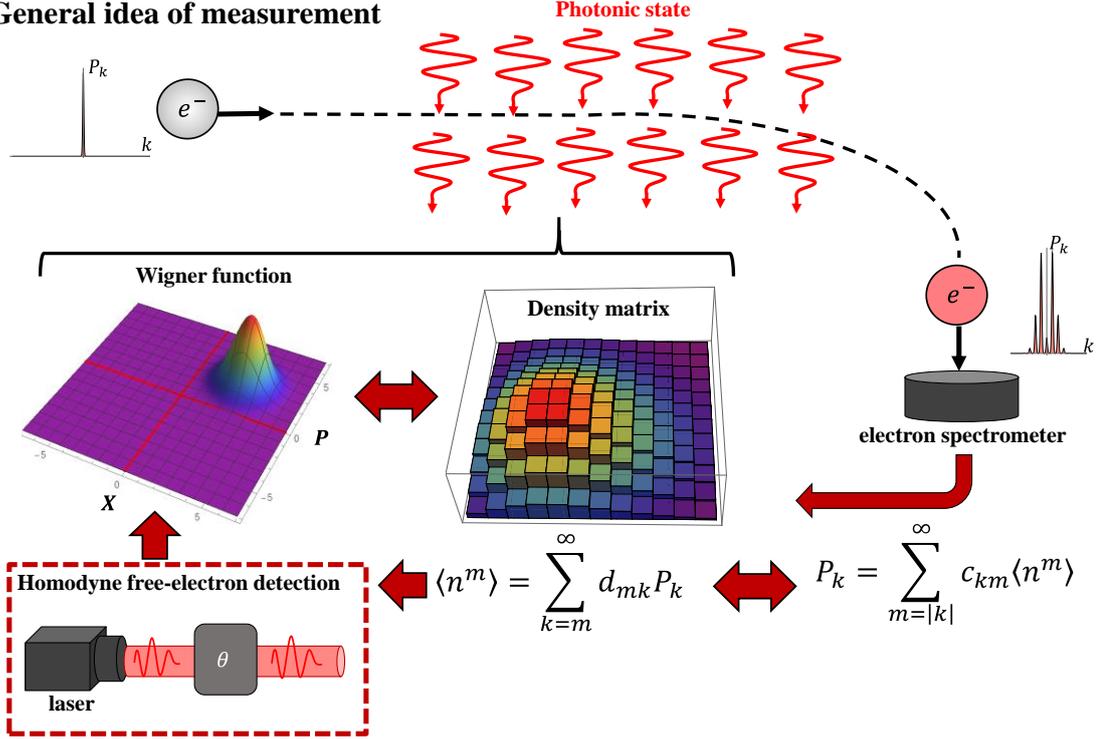

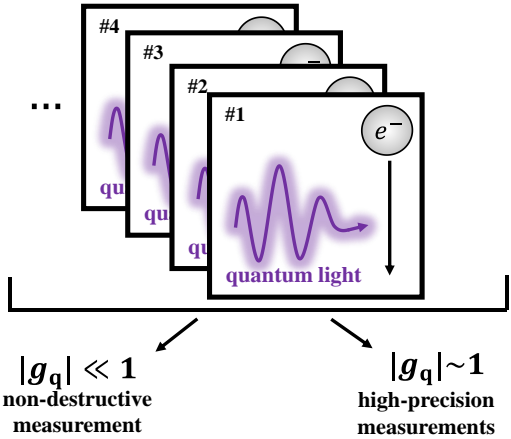

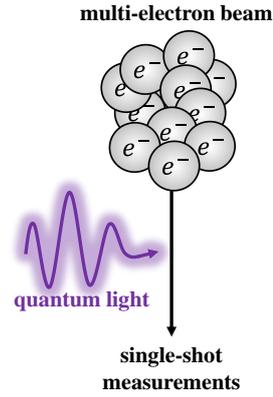

**Figure 3. Reconstruction of the quantum state of light using the measured electrons**. (a) The electron can emit/absorb a different number of photons during its interaction with the light being measured. $P_k$ is the probability to emit/absorb $k$ photons, depending on the photon number moments $\langle n^m \rangle$. The direct mathematical correspondence between $\langle n^m \rangle$ and $P_k$ (Eq. (4)) allows us to reconstruct one from a measurement of the other (Eq. (5)). We further propose free-electron homodyne detection, with which the electron energy spectrum enables reconstructing the Wigner function or, equivalently, the entire density matrix of the light (Section IV). (b) By repeating the measurement multiple times, we can reconstruct the photonic state of down to a few photons with high precision (if $|g_q| \sim 1$), or perform non-destructive measurements, where the post-interaction change in the photonic state is negligible (if $|g_q| \ll 1$). (c) Using many electrons that interact with the same pulse of light, the complete photonic state can be measured in a single shot, which is attractive for the detection of non-repetitive photonic-states.

Eq. (5) is a central result of this work. From a fundamental point of view, it shows that *a single electron* following the interaction with an optical field contains *all the information* about the photon statistics. However, all the information is hidden in the superposition coefficients of the electron's wavefunction, and it is necessary to repeat the experiment for a large number of electrons to obtain this information. The technique suggested here has some similarities to number-resolving detectors [9,15,16,56,57] – both can resolve the photon statistics of light. Despite this similarity, there are key differences: many number-resolving detectors rely on multiple single-photon detectors, each detecting one photon. In contrast, in FEQOD, a single electron can detect in principle any number of photons. The reason is that the interaction between the electron and photonic state is highly nonlinear and unbounded in its transitions, enabling the detection of a multi-photon state with just one electron.

Now let us discuss two regimes of the interaction: the strong-interaction ($|g_\mathrm{q}|\sim 1$) and weak-interaction ($|g_\mathrm{q}| \ll 1$) regimes. In the strong-interaction regime, it is possible to extract high-order photon number moments with better precision and fewer electrons. However, this comes at a price of more substantial changes in the post-interaction state of light (SM Sections VII, VIII). Thus, to measure the photon state with high precision, one needs a repetitive measurement (Fig. 3b) with the same quantum state of light at each interaction.

In the weak-interaction regime, more electrons are required to resolve high-order photon number moments $\langle n^m \rangle$. However, this ensures that the measurement is *non-destructive* (SM Sections VII, VIII, X) or, in principle, could be made in a *single-shot* (SM Section IX). Given a large number of photons so that $|g_\mathrm{q}|\sqrt{\langle n \rangle} \sim 1$ while $|g_\mathrm{q}| \ll 1$, we can use a pulse of many electrons [51,52] to estimate the quantum properties of light in one shot, i.e., using only one such pulse (Fig. 3c). Interestingly, the change of the photon state, averaged over the final electron states, equals exactly to zero in the case of Eq. (3) (SM Section VII).

In both the weak- and strong-interaction regimes, the precision of FEQOD is connected with the number of electrons – either in a single-shot scheme or in a repetitive measurement (SM Section VI). Moreover, fluctuations in the geometry of the coupling structure, the intensity of light, and many other parameters can also affect precision. All these fluctuations can be effectively considered as variations in the coupling constant $g_\mathrm{q}$. We find that the standard deviation of measured statistics depends linearly on the standard deviation of $g_\mathrm{q}$ and linearly on $m$ when we go to the higher moments: $\Delta \langle n^m \rangle / \langle n^m \rangle \approx 2m \cdot \Delta |g_\mathrm{q}|/|g_\mathrm{q}|$ (SM Section VI).

The temporal resolution of FEQOD depends on the electron pulse duration. Standard PINEM experiments [36–44] operate at a few hundreds of fs. By pre-shaping the electron, it was shown that one could reach attosecond timescales, with prospects for sub-attosecond [40,53–55]. Together with the non-destructive capabilities in the weak-interaction regime, FEQOD shows promise for advances in continuous-variable quantum information [66].

**Section IV – Full quantum state tomography and degrees of coherence**

In the previous section, we showed how the photon statistics could be extracted from the QPINEM spectrum. However, the reconstruction of the entire quantum state of light remained untreated. In this final section, we propose a method that uses FEQOD for full quantum state reconstruction and measurement of the light's second-order temporal coherence.

The FEQOD approach enables the concept of free-electron homodyne detection, analogous to homodyne quantum tomography [67–69]. In the scheme, we propose (Figure 4a), each free electron has two points of interaction with light: (1) with a strong coherent laser (acting as the local oscillator) characterized by a relative phase $\theta$ and (2) with the quantum light that should be reconstructed (theory of two-point interaction is described in SM Section I). The entire photonic state can be found by extracting the moments $\langle n^m(\theta) \rangle$ for each $\theta$ (using an optical delay) using Eq. (5). This way, we find the probability distribution of the field quadrature $X(\theta)$ (SM Section XI). The entire photonic state can then be reconstructed using the conventional homodyne quantum tomography method [69]. In many cases, the procedure is easier, for example, the full reconstruction of Gaussian states (coherent, squeezed, thermal, etc.) requires only the first two moments ($m = 1,2$). Our proposal of free-electron homodyne tomography can be non-destructive; we find that the electron interaction only causes a small perturbation to the on-diagonal and off-diagonal elements of the photonic density matrix in the weak-interaction regime (SM Section X).

Figures 4(b-e) present the free-electron homodyne tomography concept for several photonic states: plotting the Wigner functions that can be reconstructed from multiple electron energy spectra for varying $\theta$ values. To emphasize that free-electron homodyne quantum state tomography can also identify the quantum phase, we plot the difference in electron spectrum between the Schrodinger cat state (Figure 4e) and the corresponding mixed state of two coherent states (Figure 4f). Note that our scheme can be generalized to other variants by flipping the order of interactions with the laser and the quantum light. The accuracy of the suggested scheme is limited by the number of electrons involved in the measurement (SM Section VI) and fluctuations of the coupling constant $g_q$ (SM Section VI).

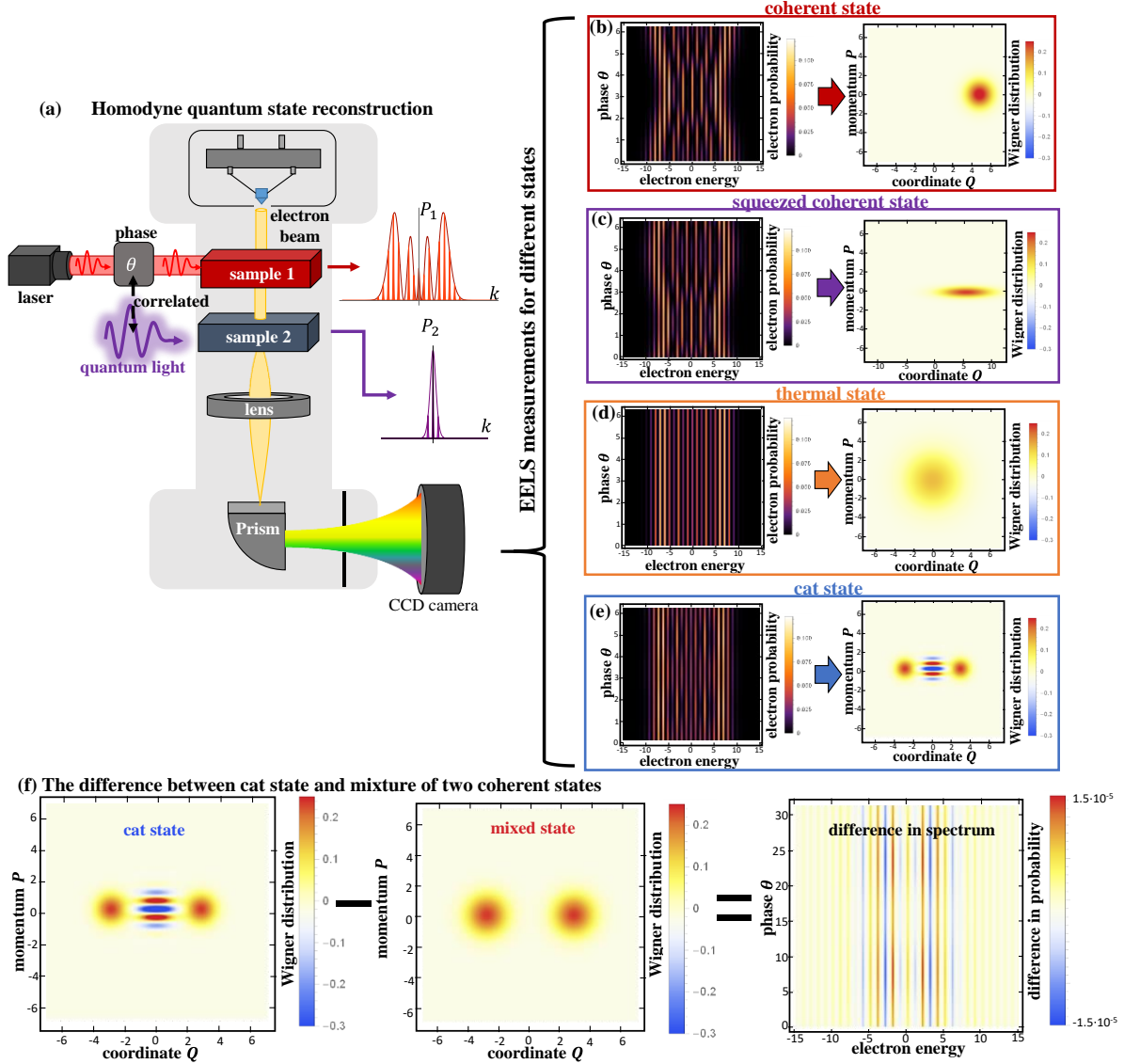

**Figure 4. Homodyne detection by free electrons for full quantum state tomography.** (a) The electron interacts with the quantum light we want to reconstruct and with an intense laser of relative phase $\theta$ (the average number of photons in the laser pulse is 100 times larger than the average number of photons in the quantum light). The reconstruction is based on measuring three different electron energy spectra: laser-only $P_1$, quantum-light-only $P_2$, and for the interaction with both sources, for different phases $\theta$, $P(\theta)$. (b-e) Examples of electron energy spectra for different photonic states, each with its corresponding Wigner function: (b) coherent state, (c) phase squeezed state, (d) thermal state, and (e) Schrodinger cat state. In the maps of electron spectra, the x-axis is the electron energy, the y-axis is the phase $\theta$, and the color denotes the probabilities. (f) The difference in electron energy spectra between a cat state $(|\alpha_1 = 2\rangle + |\alpha_2 = -2\rangle)/\sqrt{2}$ and a mixed state of two coherent states with the same coherent amplitudes.

Another significant quantum optical observable that can be measured with FEQOD is the temporal coherence of the field. Figure 5 presents a scheme for such a measurement. Our scheme utilizes two points of interaction, similar to the free-electron homodyne detection scheme. One interaction is with the quantum light that we want to measure, and the other interaction is with the same light, delayed by time $\tau$, thus resembling the HBT setup [14]. Such schemes allow for measuring modified versions of the first two degrees of quantum coherence, which we define as $\tilde{g}^{(1)}(\tau) = \frac{2\langle \tilde{n}(\tau)\rangle - \langle \tilde{n}(0)\rangle}{\langle \tilde{n}(0)\rangle}$ and $\tilde{g}^{(2)}(\tau) = 2\frac{\langle \tilde{n}^2(\tau)\rangle - \langle \tilde{n}(\tau)\rangle}{\langle \tilde{n}(\tau)\rangle^2} - \frac{\langle \tilde{n}^2(0)\rangle - \langle \tilde{n}(0)\rangle}{\langle \tilde{n}(0)\rangle^2}$, where $\langle \tilde{n}(\tau)\rangle = \langle \frac{1}{4}(a^\dagger(\tau) + a^\dagger(0))(a(\tau) + a(0))\rangle$ and $\langle \tilde{n}^2(\tau)\rangle = \langle \frac{1}{16}[(a^\dagger(\tau) + a^\dagger(0))(a(\tau) + a(0))]^2\rangle$. These temporal coherence functions differ from the conventional $g^{(1)}(\tau) = \langle a^\dagger(\tau)a(0)\rangle$ and $g^{(2)}(\tau) = \langle a^\dagger(0)a^\dagger(\tau)a(\tau)a(0)\rangle$, nevertheless, they provide the same information about the quantum light as the conventional definitions (also, in the case of zero delay $\tilde{g}^{(1,2)}(0) = g^{(1,2)}(0)$). Further discussion is provided in SM Section XII.

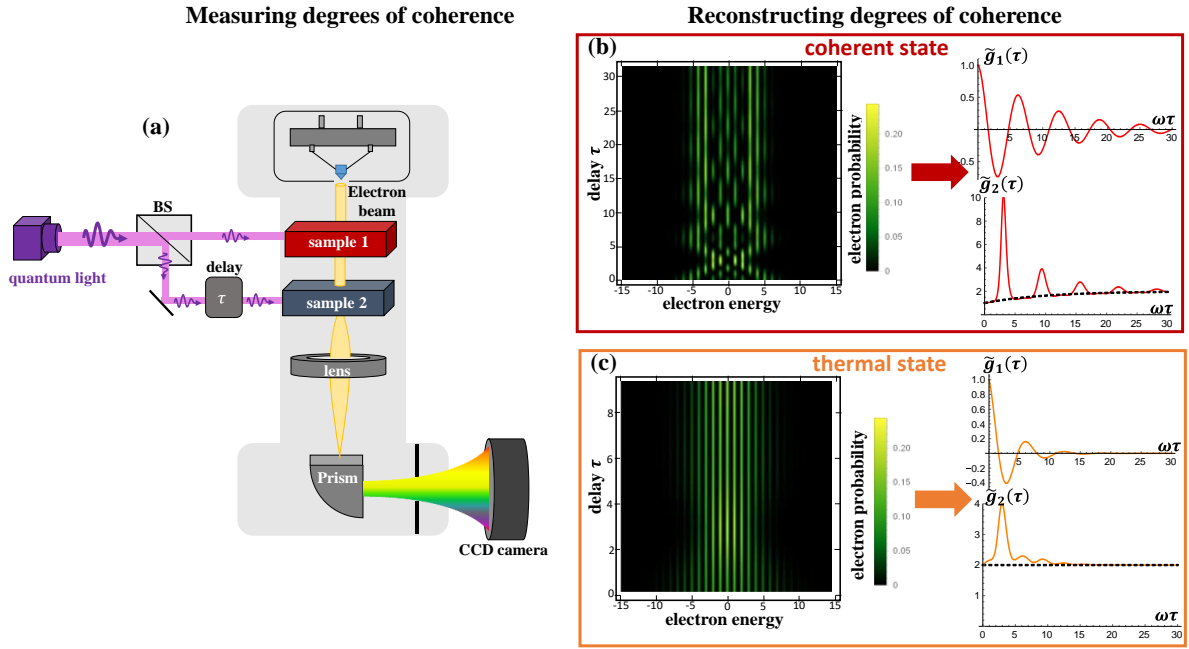

**Figure 5. Free-electron Hanbury-Brown-Twiss experiment**. **(a)** The electron interacts with the quantum light twice with a relative delay $\tau$ (presented in units of $\omega^{-1}$). We can also control the absolute delay between both light pulses and the electron pulse (not shown). The electron energy spectrum is measured for different $\tau$ values. **(b-c)** Maps of electron spectra and plots of the modified correlation functions $\tilde{g}_1(\tau)$ and $\tilde{g}_2(\tau)$ for **(b)** coherent light and **(c)** thermal light. For regular parameters in ultrafast quantum optics ($\omega/2\pi \sim 10^{14} - 10^{15}$ Hz), the resulting time resolution is a few femtoseconds, with the corresponding bandwidth reaching hundreds-to-thousands of THz. (The simulated bandwidth of the quantum light is taken as $\Delta\omega = 0.1\omega$).

**Section V – Discussion and Conclusions**

To conclude, we introduce a fundamentally new method for measuring the quantum state of light using free-electrons with ultrafast time resolution. At its core, what enables our proposed FEQOD approach and all its variants, is the fact that the electron can become entangled with the light [49], and thus after tracing-out the light, the information about its state is contained in the electron energy spectrum. From a theoretical point of view, our work generalizes previous works on QPINEM [49,63]. We derive an analytical expression for the electron's spectrum, which captures the regimes of high photon flux as well as small photon flux with arbitrary statistics and can be achieved experimentally by increasing the coupling between light and free electrons. Our analytical derivation simplifies the analysis of the PINEM interaction and is further applicable in the fast-developing field of free-electron–bound-electron interaction [70–74]. From a practical point of view, we analyzed the potential capabilities and applications of the FEQOD concept. We analyzed possible experimental setups that can measure photon statistics, reconstruct the entire quantum state, extract the temporal coherence at its different degrees, and emphasized its prospects for unprecedented temporal resolution and non-destructive measurement.

We believe that the rapid development of PINEM technology in recent years [39,41–43] makes the predictions presented in this work achievable experimentally in the near future. Specifically, recent experimental demonstrations of coupling of light and free-electrons display a large variation in the coupling constant $g_q$, mostly still in the weak-interaction regime [36,39,41]. Promising structures for reaching the strong-interaction regime include photonic crystals [41], whispering gallery modes in microspheres [43], Cherenkov phase-matched optical structures [42], and silicon photonic dielectric laser accelerators [48]. A coupling constant of $|g_q|\sim 1$, which is deep in the strong-interaction regime, was recently observed in a hybrid photonic-plasmonic structure [64], showing spontaneous emission of a few plasmonic quanta from each electron, which serves as a signature of the large $g_q$.

Existing quantum optical detectors are only rarely sensitive to the rich information contained in interactions with multi-photon quantum states of light, often requiring multiple detectors [13,33], and limited by their integration time and detection rate [75]. Novel approaches are being developed to overcome the latter limit [34,35]. In comparison, the proposed FEQOD concept is not limited by the photon number and operates at larger detection rates due to the short electron pulse duration.

Under the condition of quantum light of many photons, FEQOD enables a *single-shot* quantum-optical measurement, which can be performed using pulses of many-electrons [51,52]. This approach deviates from conventional quantum measurements [1] that require to reproduce the photonic state and repeat the measurement with different realizations of the photonic state until sufficient signal is acquired. In contrast, the *single-shot* scheme enables to measure the photon statistics even for a non-reproducible photonic state. This capability paves the way to separating between classical and quantum uncertainties in quantum light.

Altogether, the proposed concept of FEQOD links the fast-developing field of PINEM and the well-established field of quantum optics. FEQOD adds a unique capability to the toolbox of quantum optics by combining ultrafast detection and non-destructive measurements. Future progress in free-electron quantum optics could have an impact in numerous areas, including quantum tomography [67–69], weak measurements [76–78], non-destructive measurements, and quantum imaging [79,80].

**References**

1. Scully, M. O. & S., Z. M. *Quantum Optics*. (Cambridge University press, 1997).

2. Garrison, J. & Chiao, R. *Quantum Optics*. *Quantum Optics* vol. 9780198508861 (Oxford University Press, 2008).

3. Aspect, A., Dalibard, J. & Roger, G. Experimental test of Bell's inequalities using time- varying analyzers. *Phys. Rev. Lett.* **49**, 1804–1807 (1982).

4. Hong, C. K., Ou, Z. Y. & Mandel, L. Measurement of subpicosecond time intervals between two photons by interference. *Phys. Rev. Lett.* **59**, 2044–2046 (1987).

5. Grangier, P., Roger, G. & Aspect, A. Experimental evidence for a photon anticorrelation effect on a beam splitter: A new light on single-photon interferences. *EPL* **1**, 173–179 (1986).

6. Gao, W. B. *et al.* Experimental demonstration of a hyper-entangled ten-qubit Schrödinger cat state. *Nat. Phys.* **6**, 331–335 (2010).

7. Vahlbruch, H., Mehmet, M., Danzmann, K. & Schnabel, R. Detection of 15 dB Squeezed States of Light and their Application for the Absolute Calibration of Photoelectric Quantum Efficiency. *Phys. Rev. Lett.* **117**, 110801 (2016).

8. Kok, P. *et al.* Linear optical quantum computing with photonic qubits. *Rev. Mod. Phys.* **79**, 135–174 (2007).

9. Politi, A., Matthews, J. C. F. & O'Brien, J. L. Shor's quantum factoring algorithm on a photonic chip. *Science* vol. 325 1221 (2009).

10. Jouguet, P., Kunz-Jacques, S., Leverrier, A., Grangier, P. & Diamanti, E. Experimental demonstration of long-distance continuous-variable quantum key distribution. *Nat. Photonics* **7**, 378–381 (2013).

11. Aasi, J. *et al.* Enhanced sensitivity of the LIGO gravitational wave detector by using squeezed states of light. *Nat. Photonics* **7**, 613–619 (2013).

12. Boaron, A. *et al.* Secure Quantum Key Distribution over 421 km of Optical Fiber. *Phys. Rev. Lett.* **121**, 190502 (2018).

13. Zhong, H.-S. *et al.* Quantum computational advantage using photons. *Science (80-. ).* (2020).

14. Brown, R. H. & Twiss, R. Q. Correlation between Photons in two Coherent Beams of Light. *Nature* **177**, 27–29 (1956).

15. Jiang, L. A., Dauler, E. A. & Chang, J. T. Photon-number-resolving detector with 10 bits of resolution. *Phys. Rev. A - At. Mol. Opt. Phys.* **75**, 062325 (2007).

16. Divochiy, A. *et al.* Superconducting nanowire photon-number-resolving detector at telecommunication wavelengths. *Nat. Photonics* **2**, 302–306 (2008).

17. Cooper, M., Söller, C. & Smith, B. J. High-stability time-domain balanced homodyne detector for ultrafast optical pulse applications. in *Optics InfoBase Conference Papers* QTu3E.5 (Optical Society of America, 2012).

18. Zhang, G. *et al.* An integrated silicon photonic chip platform for continuous-variable quantum key distribution. *Nat. Photonics* **13**, 839–842 (2019).

19. Leonhardt, U. *Measuring the quantum state of light*. vol. 22 (Cambridge university press, 1997).

20. Lvovsky, A. I. & Raymer, M. G. Continuous-variable optical quantum-state tomography. *Rev. Mod. Phys.* **81**, 299–332 (2009).

21. Bent, N. *et al.* Experimental realization of quantum tomography of photonic qudits via symmetric informationally complete positive operator-valued measures. *Phys. Rev. X* **5**, 41006 (2015).

22. Ahn, D. *et al.* Adaptive compressive tomography with no a priori information. *Phys. Rev. Lett.* **122**, 100404 (2019).

23. Nehra, R. *et al.* State-independent quantum state tomography by photon-number-resolving measurements. *Optica* **6**, 1356–1360 (2019).

24. Pellegrini, S. *et al.* Design and performance of an InGaAs-InP single-photon avalanche diode detector. *IEEE J. Quantum Electron.* **42**, 397–403 (2006).

25. Nehra, R., Chang, C.-H., Yu, Q., Beling, A. & Pfister, O. Photon-number-resolving segmented detectors based on single-photon avalanche-photodiodes. *Opt. Express* **28**, 3660–3675 (2020).

26. Natarajan, C. M., Tanner, M. G. & Hadfield, R. H. Superconducting nanowire single-photon detectors: Physics and applications. *Superconductor Science and Technology* vol. 25 063001 (2012).

27. Becker, W. *Advanced Time-Correlated Single Photon Counting Techniques*. vol. 81 (Springer Berlin Heidelberg, 2005).

28. Raimond, J. M., Brune, M. & Haroche, S. Colloquium: Manipulating quantum entanglement with atoms and photons in a cavity. *Rev. Mod. Phys.* **73**, 565–582 (2001).

29. Brune, M. *et al.* Quantum rabi oscillation: A direct test of field quantization in a cavity. *Phys. Rev. Lett.* **76**, 1800–1803 (1996).

30. Goy, P., Raimond, J. M., Gross, M. & Haroche, S. Observation of cavity-enhanced single-atom spontaneous emission. *Phys. Rev. Lett.* **50**, 1903–1906 (1983).

31. Herkommer, A. M., Akulin, V. M. & Schleich, W. P. Quantum demolition measurement of photon statistics by atomic beam deflection. *Phys. Rev. Lett.* **69**, 3298–3301 (1992).

32. Khosa, A. H. & Zubairy, M. S. Measurement of the Wigner function via atomic beam deflection in the Raman-Nath regime. *J. Phys. B At. Mol. Opt. Phys.* **39**, 5079–5089 (2006).

33. Hübel, H. *et al.* Direct generation of photon triplets using cascaded photon-pair sources. *Nature* **466**, 601–603 (2010).

34. Shaked, Y. *et al.* Lifting the bandwidth limit of optical homodyne measurement with broadband parametric amplification. *Nat. Commun.* **9**, 609 (2018).

35. Frascella, G. *et al.* Wide-field SU(1,1) interferometer. *Optica* **6**, 1233–1236 (2019).

36. Barwick, B., Flannigan, D. J. & Zewail, A. H. Photon-induced near-field electron microscopy. *Nature* **462**, 902–906 (2009).

37. Park, S. T., Lin, M. & Zewail, A. H. Photon-induced near-field electron microscopy (PINEM): Theoretical and experimental. *New J. Phys.* **12**, 123028 (2010).

38. García De Abajo, F. J. Optical excitations in electron microscopy. *Reviews of Modern Physics* vol. 82 209–275 (2010).

39. Feist, A. *et al.* Quantum coherent optical phase modulation in an ultrafast transmission electron microscope. *Nature* **521**, 200–203 (2015).

40. Priebe, K. E. *et al.* Attosecond electron pulse trains and quantum state reconstruction in ultrafast transmission electron microscopy. *Nat. Photonics* **11**, 793–797 (2017).

41. Wang, K. *et al.* Coherent interaction between free electrons and a photonic cavity. *Nature* **582**, (2020).

42. Dahan, R. *et al.* Resonant phase-matching between a light wave and a free-electron wavefunction. *Nat. Phys.* **16**, 1123–1131 (2020).

43. Kfir, O. *et al.* Controlling free electrons with optical whispering-gallery modes. *Nature* **582**, 46–49 (2020).

44. Rivera, N. & Kaminer, I. Light–matter interactions with photonic quasiparticles. *Nat. Rev. Phys.* **2**, 538–561 (2020).

45. Mangles, S. P. D. *et al.* Monoenergetic beams of relativistic electrons from intense laser–plasma interactions. *Nature* **431**, 535–538 (2004).

46. Norreys, P. A. Laser-driven particle acceleration. *Nat. Photonics* **3**, 423–425 (2009).

47. England, R. J. *et al.* Dielectric laser accelerators. *Rev. Mod. Phys.* **86**, 1337–1389 (2014).

48. Adiv, Y. *et al.* Observation of the Quantum Nature of Laser-Driven Particle Acceleration. in *Conference Proceedings - Lasers and Electro-Optics Society Annual Meeting-LEOS* vols 2020-May (Institute of Electrical and Electronics Engineers Inc., 2020).

49. Kfir, O. Entanglements of Electrons and Cavity Photons in the Strong-Coupling Regime. *Phys. Rev. Lett.* **123**, 103602 (2019).

50. Goldberg, A. Z., Klimov, A. B., Grassl, M., Leuchs, G. & Sánchez-Soto, L. L. Extremal quantum states. *AVS Quantum Sci.* **2**, 44701 (2020).

51. Gahlmann, A., Tae Park, S. & Zewail, A. H. Ultrashort electron pulses for diffraction, crystallography and microscopy: Theoretical and experimental resolutions. *Phys. Chem. Chem. Phys.* **10**, 2894–2909 (2008).

52. Schroeder, C. B., Lee, P. B., Wurtele, J. S., Esarey, E. & Leemans, W. P. Generation of ultrashort electron bunches by colliding laser pulses. *Phys. Rev. E - Stat. Physics, Plasmas, Fluids, Relat. Interdiscip. Top.* **59**, 6037–6047 (1999).

53. Morimoto, Y. & Baum, P. Diffraction and microscopy with attosecond electron pulse trains. *Nat. Phys.* **14**, 252–256 (2018).

54. Kozák, M., Schönenberger, N. & Hommelhoff, P. Ponderomotive Generation and Detection of Attosecond Free-Electron Pulse Trains. *Phys. Rev. Lett.* **120**, 103203 (2018).

55. Vanacore, G. M. *et al.* Attosecond coherent control of free-electron wave functions using semi-infinite light fields. *Nat. Commun.* **9**, 2694 (2018).

56. Fitch, M. J., Jacobs, B. C., Pittman, T. B. & Franson, J. D. Photon-number resolution using time-multiplexed single-photon detectors. *Phys. Rev. A - At. Mol. Opt. Phys.* **68**, 6 (2003).

57. Marsili, F. *et al.* Physics and application of photon number resolving detectors based on superconducting parallel nanowires. *New J. Phys.* **11**, 45022 (2009).

58. Polman, A., Kociak, M. & García de Abajo, F. J. Electron-beam spectroscopy for nanophotonics. *Nat. Mater.* **18**, 1158–1171 (2019).

59. Meuret, S. *et al.* Photon Bunching in Cathodoluminescence. *Phys. Rev. Lett.* **114**, 197401 (2015).

60. Tizei, L. H. G. & Kociak, M. Spatially Resolved Quantum Nano-Optics of Single Photons Using an Electron Microscope. *Phys. Rev. Lett.* **110**, 153604 (2013).

61. Kociak, M. & Zagonel, L. F. Cathodoluminescence in the scanning transmission electron microscope. *Ultramicroscopy* **176**, 112–131 (2017).

62. Hayun, A. Ben *et al.* Shaping Quantum Photonic States Using Free Electrons. (2020).

63. Giulio, V. Di, Kociak, M. & de Abajo, F. J. G. Probing quantum optical excitations with fast electrons. *Optica* **6**, 1524–1534 (2019).

64. Adiv, Y. *et al.* Observing two-dimensional Cherenkov radiation and its quantized nature. *Prep.*

65. Garcia De Abajo, F. J., Asenjo-Garcia, A. & Kociak, M. Multiphoton absorption and emission by interaction of swift electrons with evanescent light fields. *Nano Lett.* **10**, 1859–1863 (2010).

66. Weedbrook, C. *et al.* Gaussian quantum information. *Rev. Mod. Phys.* **84**, 621–669 (2012).

67. *Quantum State Estimation*. vol. 649 (Springer Berlin Heidelberg, 2004).

68. Altepeter, J. B., Jeffrey, E. R. & Kwiat, P. G. Photonic State Tomography. in (eds. Berman, P. R. & Lin, C. C.) vol. 52 105–159 (Academic Press, 2005).

69. Leonhardt, U. & Paul, H. Measuring the quantum state of light. *Progress in Quantum Electronics* vol. 19 89–130 (1995).

70. Gover, A. & Yariv, A. Free-Electron-Bound-Electron Resonant Interaction. *Phys. Rev. Lett.* **124**, 064801 (2020).

71. Zhao, Z., Sun, X.-Q. & Fan, S. Quantum entanglement and modulation enhancement of free-electron-bound-electron interaction. (2020).

72. Ruimy, R., Gorlach, A., Mechel, C., Rivera, N. & Kaminer, I. Towards atomic-resolution quantum measurements with coherently-shaped free electrons. (2020).

73. Gover, A. *et al.* Resonant Interaction of Modulation-correlated Quantum Electron Wavepackets with Bound Electron States. (2020).

74. de Abajo, F. J. G. & Di Giulio, V. Quantum and Classical Effects in Sample Excitations by Electron Beams. (2020).

75. Hadfield, R. H. Single-photon detectors for optical quantum information applications. *Nat. Photonics* **3**, 696–705 (2009).

76. Aharonov, Y., Albert, D. Z. & Vaidman, L. How the result of a measurement of a component of the spin of a spin-1/2 particle can turn out to be 100. *Phys. Rev. Lett.* **60**, 1351–1354 (1988).

77. Hosten, O. & Kwiat, P. Observation of the Spin Hall Effect of Light via Weak Measurements. *Science (80-. ).* **319**, 787–790 (2008).

78. Dixon, P. Ben, Starling, D. J., Jordan, A. N. & Howell, J. C. Ultrasensitive Beam Deflection Measurement via Interferometric Weak Value Amplification. *Phys. Rev. Lett.* **102**, 173601 (2009).

79. Brida, G., Genovese, M. & Ruo Berchera, I. Experimental realization of sub-shot-noise quantum imaging. *Nat. Photonics* **4**, 227–230 (2010).

80. Genovese, M. Real applications of quantum imaging. *J. Opt.* **18**, 73002 (2016).

# Ultrafast non-destructive measurement of the quantum state of light with free electrons

# Supplementary Material

**Alexey Gorlach[1], Aviv Karnieli[2], Raphael Dahan[1], Eliahu Cohen[3], Avi Pe'er[3], and Ido Kaminer[1]**

1. Solid State Institute, Technion-Israel Institute of Technology, Haifa 32000, Israel
2. Sackler School of Physics, Faculty of Exact Sciences, Tel Aviv University, Tel Aviv, 69978 Israel.
3. Department of Physics and BINA Center of Nanotechnology, Bar-Ilan University, 52900 Ramat-Gan, Israel

kaminer@technion.ac.il

**Section I – Generalizing the electron-photon interaction to multiple independent interactions**

This section shows that a few consecutive interactions of an electron with monochromatic light can be described as one interaction. Let us start from the scattering matrix for one interaction point (Eq. (1) in the main text):

$$S = \exp[gba^\dagger - g^*b^\dagger a], \tag{1}$$

where $g$ is a quantum coupling constant; $b$ and $b^\dagger$ are the ladder operators for the free electron; $a$ and $a^\dagger$ are ladder operators for the nearfield light, the commutation relation for these operators is the following: $[a, a^\dagger] = 1$. Let us consider $N$ interaction points with independent modes of light:

$$S = \exp\left[gb\sum_{i=1}^{N} a_i^\dagger - g^*b^\dagger \sum_{i=1}^{N} a_i\right], \tag{2}$$

where for different operators $a_i$ ($i$ denotes different, spatially-separated points of interaction) the following conditions are satisfied: $[a_j, a_i^\dagger] = \delta_{ij}$, $[a_i, a_j] = 0$ (i.e., no spatial overlap exists between the modes at different interaction points). In this case, we introduce new notions:

$$\begin{cases} \tilde{a} \equiv \dfrac{1}{\sqrt{N}} \sum_{i=1}^{N} a_i, \\ \tilde{g} \equiv \sqrt{N} g \end{cases} \tag{3}$$

We get Eq. (1), but with new operators $\tilde{a}$ and $\tilde{a}^\dagger$ such that the commutation relation $[\tilde{a}, \tilde{a}^\dagger] = 1$ still holds and with a new coupling constant defined by Eq. (3). Hence, we show that the problem of many independent points of interaction can be reduced to the problem of one point of interaction but with a different definition of the coupling constant $g$ and creation operators $a, a^\dagger$.

**Section II – The exact quantum theory of PINEM and an approximation for weak coupling and a large number of photons**

In this section, we find the probability $P_{nk}$ to have $n$ photons and final electron energy of $E_0 + k\hbar\omega$, where $E_0$ is the initial electron energy, and $\hbar\omega$ is the energy of the nearfield photons interacting with the electron. We derive an exact solution to this problem for the first time and show under what assumptions it gives an already known approximated solution [1, 2].

We assume that the initial electron is monoenergetic, i.e., it is found in a single energy eigenstate of the eigenvalue $E_0$. We denote such a state as $|E_0\rangle$. We assume that the electron interacts with an arbitrary initial state of light, which can be described by the density matrix $\rho_{\text{ph}}$. Then, the initial joint state of the electron-light system is: $|\Psi_i\rangle = |E_0\rangle\rho_{\text{ph}}$. The state of the system after the interaction is given by:

$$|\Psi_f\rangle = S|\Psi_i\rangle, \tag{4}$$

where $S$ is given by Eq. (1). The probability $P_{nk}$ equals to:

$$P_{nk} = \langle n|S_k \rho_{\text{ph}} S_k^\dagger|n\rangle, \tag{5}$$

where $|n\rangle$ is the Fock state with $n$ photons, $S_k = \langle E_k|S|E_0\rangle$ and $\rho_{\text{ph}}$ is the initial density matrix of the photonic state defined by $\rho_{\text{ph}} = |\psi_{\text{ph}}\rangle\langle\psi_{\text{ph}}|$, or more generally, as any mixed initial state. Let us calculate $S_k$:

$$S_k = \langle E_k| \exp[g_q ba^\dagger - g_q^* b^\dagger a] |E_0\rangle = e^{\frac{|g|^2}{2}} \langle E_k| \sum_{m,l=0}^{\infty} \frac{(-g^*)^m g^l}{m!\, l!} (b^\dagger a)^m (ba^\dagger)^l |E_0\rangle.$$

Now we should consider two different cases $k > 0$ and $k < 0$.

*The case of $k > 0$:*

In this case, only terms that satisfy $m = k + l$ are non-zero:

$$S_k = (-g^*)^k e^{\frac{|g|^2}{2}} a^k \sum_{l=0}^{\infty} \frac{(-1)^l |g|^{2l}}{(l+k)!\, l!} (a)^l (a^\dagger)^l.$$

If we introduce the photon number operator $\hat{n} = a^\dagger a$, then:

$$a^l a^{\dagger l} = \frac{(\hat{n} + l)!}{\hat{n}!}. \tag{6}$$

Using Eq. (6), we get:

$$S_k = (-g^*)^k e^{\frac{|g|^2}{2}} a^k \sum_{l=0}^{\infty} \frac{(-1)^l |g|^{2l}}{(l+k)! \, l!} \frac{(\hat{n}+l)!}{\hat{n}!}.$$

Now we calculate the sum, and we get:

$$S_k = (-g^*)^k e^{\frac{|g|^2}{2}} a^k \frac{{}_1F_1(\hat{n}+1, k+1, -|g|^2)}{k!}, \tag{7}$$

where ${}_1F_1(\hat{n}+1, k+1, -|g|^2)$ is the hypergeometrical function. Thus, according to Eq. (5), the probability equals to:

$$P_{nk} = \langle n | S_k \rho_{\text{ph}} S_k^\dagger | n \rangle =$$

$$= |g|^{2k} e^{|g|^2} \frac{(n+k)!}{n!} \left| \frac{{}_1F_1(n+k+1, k+1, -|g|^2)}{k!} \right|^2 \langle n+k | \rho_{\text{ph}} | n+k \rangle =$$

$$= |g|^{2k} e^{|g|^2} \frac{(n+k)!}{n!} \left| \frac{{}_1F_1(n+k+1, k+1, -|g|^2)}{k!} \right|^2 p_{n+k},$$

where $p_n = \langle n | \rho_{\text{ph}} | n \rangle$.

*Case of $k < 0$:*

Then, only terms that satisfy $l = m + |k|$ are non-zero:

$$S_k = g^{|k|} e^{\frac{|g|^2}{2}} \left( \sum_{m=0}^{\infty} \frac{(-1)^m |g|^{2m}}{m! \, (m+|k|)!} a^m (a^\dagger)^m \right) (a^\dagger)^{|k|}.$$

Now we calculate the sum, and we get:

$$S_k = g^{|k|} e^{\frac{|g|^2}{2}} \frac{{}_1F_1(\hat{n}+1, |k|+1, -|g|^2)}{|k|!} (a^\dagger)^{|k|}, \tag{9}$$

Thus, according to Eq. (5), the probability equals to:

$$P_{nk} = |g|^{2|k|} e^{|g|^2} \left| \frac{{}_1F_1(n+1, |k|+1, -|g|^2)}{|k|!} \right|^2 \langle n | (a^\dagger)^{|k|} \rho_{\text{ph}} a^{|k|} | n \rangle =$$

$$= |g|^{2|k|} e^{|g|^2} \frac{n!}{(|k|\,)^2 (n-|k|)!} |{}_1F_1(n+1, |k|+1, -|g|^2)|^2 p_{n-|k|}.$$

where $p_n = \langle n | \rho_{\text{ph}} | n \rangle$.

Combining both results, we get:

$$P_{nk} = \begin{cases} |g|^{2k} e^{|g|^2} \dfrac{(n+k)!}{(k!)^2 n!} |_1F_1(n+k+1, k+1, -|g|^2)|^2 p_{n+k}, & k > 0 \\ |g|^{2|k|} e^{|g|^2} \dfrac{n!}{(k!)^2 (n-|k|)!} |_1F_1(n+1, |k|+1, -|g|^2)|^2 p_{n-|k|}, & k < 0 \end{cases}, \quad (10)$$

Eq. (10), as a complete analytical solution of the QPINEM problem, provides all the tools to analyze any phenomena connected with the scattering matrix defined by Eq. (1), which is more general than classical and semiclassical PINEM and may also be generalized to describe the quantum interaction between a two-level system and a free electron, that was studied recently in Refs. [3, 4].

Let us now analyze how to recover from Eq. (10) the conventional approximated result of QPINEM [1, 2]. For this, we need to take the limit $|g| \to 0, n \to \infty$:

$$\frac{{}_1F_1(n+1, k+1, -|g|^2)}{k!} \approx \sum_{l=0}^{\infty} \frac{(-1)^l (|g|\sqrt{n})^{2l}}{(l+k)!\, l!} = (|g|\sqrt{n})^{-k} J_k(2|g|\sqrt{n}),$$

where $J_k(2|g|\sqrt{n})$ is the Bessel function.

Then under the assumption of $|g| \to 0, n \to \infty$, Eq. (10) can be simplified:

$$P_{nk} \approx |J_k(2|g|\sqrt{n})|^2 p_n. \quad (11)$$

Eq. (11) is the approximation of Eq. (10) and is the conventional equation [1, 2] to describe the quantum theory of PINEM.

## Section III – The electron energy spectrum after a quantum PINEM interaction

The electron spectrum is measured after the PINEM interaction using electron energy loss spectroscopy (EELS). The formalism we build enables us to find the spectrum of the post-interaction electron. This section finds the exact solution for the EELS spectrum and compares this with an approximated conventional solution [1, 2]. Here, we generalize the electron spectrum's theoretical prediction to take into account the interaction with arbitrary quantum light, which is generally described by the density matrix.

The electron spectrum can be found from Eq. (10) by summing over all possible photon numbers in the photonic field:

$$P_k = \sum_{n=0}^{\infty} P_{nk} = \begin{cases} |g|^{2k} e^{|g|^2} \sum_{n=0}^{\infty} \frac{n!}{(k!)^2 (n-k)!} |{}_1F_1(n+1, k+1, -|g|^2)|^2 p_n, & k > 0 \\ |g|^{2|k|} e^{|g|^2} \sum_{n=0}^{\infty} \frac{(n+|k|)!}{(k!)^2 n!} |{}_1F_1(n+|k|+1, |k|+1, -|g|^2)|^2 p_n, & k < 0 \end{cases},$$

where we assume that $\frac{1}{(n-k)!} = 0$ for $k > n$. We should notice that $\sum_n^{\infty} \ldots \cdot p_n$ is averaging over all possible number of photons, hence in the most general case, we can write:

$$P_k = \begin{cases} \mathrm{Tr}\left[\rho_{\mathrm{ph}} \cdot |g|^{2k} e^{|g|^2} \frac{\hat{n}!}{(k!)^2 (\hat{n}-k)!} |{}_1F_1(\hat{n}+1, k+1, -|g|^2)|^2\right], & k > 0 \\ \mathrm{Tr}\left[\rho_{\mathrm{ph}} \cdot |g|^{2|k|} e^{|g|^2} \frac{(\hat{n}+|k|)!}{(|k|!)^2 \hat{n}!} |{}_1F_1(\hat{n}+|k|+1, |k|+1, -|g|^2)|^2\right], & k < 0 \end{cases}, \quad (12)$$

where $\rho_{\mathrm{ph}}$ is the density matrix of the initial photonic state. Eq. (12) gives the exact and the most general description of the electron's spectrum after a PINEM interaction. It corresponds to Eq. (2) in the manuscript. We can find the approximated solution using Eq. (11):

$$P_k = \sum_{n=0}^{\infty} P_{nk} = \sum_{n=0}^{\infty} |J_k(2|g|\sqrt{n})|^2 p_n. \quad (13)$$

Eq. (13) corresponds to Eq. (3) in the main text.

We now compare Eq. (12) and Eq. (13). Let us find the range of parameters where Eq. (13) provides a reasonable approximation to the electron spectrum. With this purpose, we consider the most practical example of a coherent state $|\alpha\rangle$ (e.g., as produced by laser light):

$$p_n = e^{-|\alpha|^2} \frac{|\alpha|^{2n}}{n!},$$

where $p_n$ is the portability for $n$ photons in the coherent state $|\alpha\rangle$. Fig. 1 represents the deviation of Eq. (13) from the exact solution Eq. (12) for a fixed value of $g\sqrt{\langle n \rangle} = 3$ but for different coupling constants $g$. We see a substantial deviation of Eq. (13) from Eq. (12) for $g > 0.3$, i.e., in the strong coupling regime.

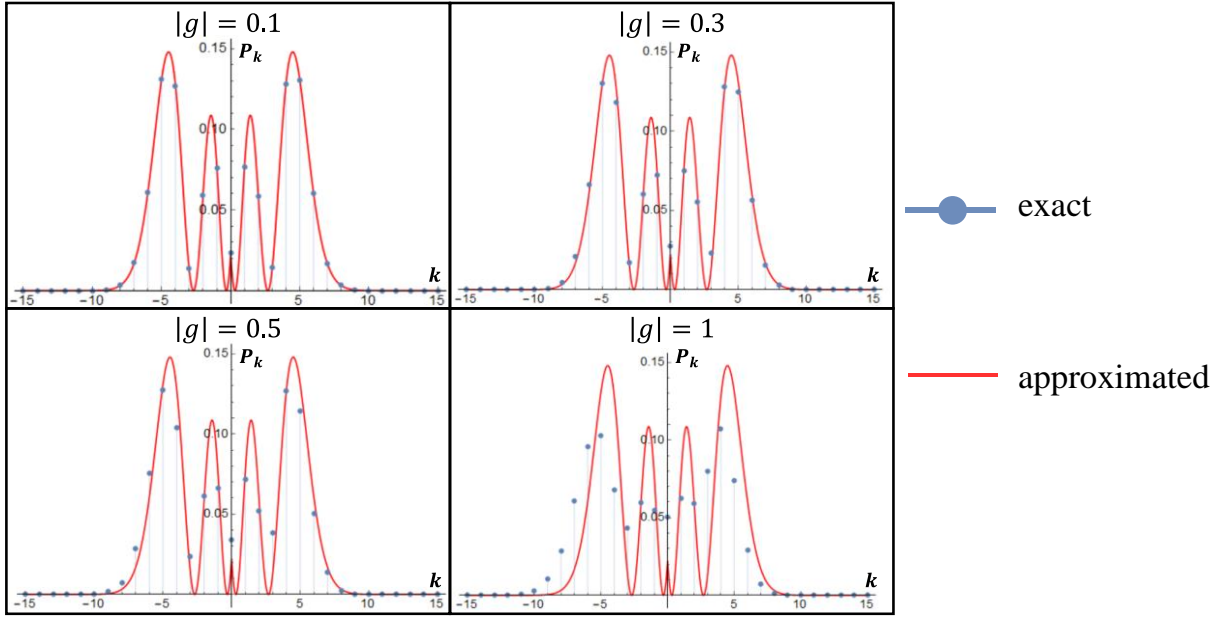

**Fig 1. Validating the approximation numerically. Comparison between the exact and the approximated electron energy spectra after the PINEM interaction**. The spectrum $P_k$ is built according to Eq. (12) (blue dots) and Eq. (13) (red curve) for the coherent state $|\alpha\rangle$. We take $|g|\sqrt{\langle n \rangle} = 3$ in all the plots, and vary the value of $|g|$. We can see that the approximate solution is very accurate for $g \leq 0.3$.

**Section IV – Approximated analytical solutions for specific quantum photonic states**

In this section, we find approximated analytical solutions to different quantum states using Eq. (13):

$$P_k = \sum_{n=0}^{\infty} |J_k(2|g|\sqrt{n})|^2 p_n.$$

These analytical solutions are a good approximation in the limit of $|g| \to 0$ and $|g|\sqrt{\langle n \rangle} \to$ const. In this limit, we can define the classical PINEM constant $\beta = |g|\sqrt{\langle n \rangle}$ as in [1]. Here we consider coherent, thermal, squeezed, and Fock states.

*Fock state*

Fock state $|n\rangle$ is defined by the following statistics:

$$\begin{cases} p_{n=\langle n \rangle} = 1 \\ p_{n \neq \langle n \rangle} = 0 \end{cases}.$$

According to Eq. (13), we get:

$$P_k = |J_k(2\beta)|^2, \tag{14}$$

where $\beta = |g|\sqrt{\langle n \rangle}$ is the strength of PINEM interaction.

*Coherent state*

The Poisson distribution describes the photon statistics of the coherent state:

$$p_n = e^{-\langle n \rangle} \frac{\langle n \rangle^n}{n!} \approx \frac{1}{\sqrt{2\pi \bar{n}}} \exp\left[-\frac{(n-\langle n \rangle)^2}{2\langle n \rangle}\right].$$

In the continuous limit, when $n \to \infty$, we get:

$$p(x)dx \approx \frac{1}{\sqrt{2\pi\langle n \rangle}} \exp\left[-\frac{(x-\langle n \rangle)^2}{2\langle n \rangle}\right] dx.$$

Then, the electron's spectrum can be written in the integral:

$$P_k \approx \int_{-\infty}^{\infty} \frac{1}{\sqrt{2\pi\langle n \rangle}} \exp\left[-\frac{x^2}{2\langle n \rangle}\right] J_{|k|}^2 \left(2|g|\sqrt{(x+\langle n \rangle)}\right) dx.$$

Up to the 2$^{nd}$ order in $\frac{x}{\langle n \rangle}$, the probability equals to:

$$P_k \approx |J_k(2\beta)|^2 + \frac{F(\beta)}{\langle n \rangle}, \tag{15}$$

where

$$F = |g|^2 \langle n \rangle |J_{|k|+1}(2\beta)|^2 + (k(k-1) - 2\beta^2)|J_{|k|}(2\beta)|^2 - \beta(k-1)J_{|k|}(2\beta)J_{|k|+1}(2\beta).$$

We should note that according to Eq. (15) coherent state in the limit of $\langle n \rangle \to \infty$ coincide with the spectrum for the Fock state Eq. (14). However, the small correction $\frac{F(\beta)}{\langle n \rangle}$ in the electron spectrum for the coherent state shows the fundamental difference between the Fock and coherent state results.

*Thermal state*

The photon statistics for the thermal state is the following:

$$p_n = \frac{1}{\langle n \rangle + 1} \left( \frac{\langle n \rangle}{\langle n \rangle + 1} \right)^n.$$

In the continuous limit, when $n \to \infty$, we get:

$$p(x)dx \approx \frac{1}{\langle n \rangle} e^{-\frac{x}{\langle n \rangle}} dx.$$

Then, the electron's spectrum can be found using the following integral:

$$P_k \approx \frac{1}{\langle n \rangle} \int_0^\infty J_{|k|}^2(2|g|\sqrt{x}) e^{-\frac{x}{\langle n \rangle}} dx.$$

After the calculation of this integral, we get:

$$P_k = e^{-2\beta^2} I_{|k|}(2\beta^2), \qquad (16)$$

where $I_k$ is the modified Bessel function of the first time.

*Squeezed vacuum state*

The statistics of the squeezed vacuum state has the following form:

$$\begin{cases} p_{2n+1} = 0 \\ p_{2n} = \frac{1}{\sqrt{1 + \langle n \rangle}} \left( \frac{\langle n \rangle}{\langle n \rangle + 1} \right)^{2n} \frac{(2n)!}{2^{2n}(n!)^2} \end{cases}.$$

In the continuous limit, it gives:

$$p(x)dx = \frac{1}{\sqrt{2\pi \langle n \rangle x}} e^{-\frac{x}{2\langle n \rangle}} dx.$$

This continuous distribution leads to the following answer:

$$P_k = \frac{2^{|k|}(\beta)^{2|k|}}{\sqrt{\pi}} \frac{\Gamma\left[\frac{1}{2}+|k|\right]}{(|k|!)^2} \, {}_2F_2\left(\left\{\frac{1}{2}+|k|,\frac{1}{2}+|k|\right\},\{1+|k|,1+2|k|\},-8\beta^2\right), \qquad (17)$$

where $\Gamma$ is gamma-function, ${}_2F_2$ is a hypergeometric function.

*Squeezed coherent state*

The squeezed coherent state is described by:

$$p_n = (1-|z|^2)^{1/2} \frac{|z|^n}{2^n n!} \exp\left[-|\alpha|^2 - \frac{1}{2}(\alpha^2 z^* + \alpha^{*2} z)\right] \cdot \left|H_n\left[\frac{(\alpha+z\alpha^*)}{\sqrt{2z}}\right]\right|^2,$$

where $H_n$ is the Hermite polynomial; $\alpha$ is the coherent amplitude; $z = e^{2i\phi}\tanh|r|$, where $|r|$ is the squeezing parameter, and $\phi$ is the squeezing phase ($\phi = 0$ and $\phi = \pi/2$ correspond to amplitude squeezed and phase squeezed light, respectively). Here we did not derive an analytical approximation of Eq. (13) for the spectrum.

All the results are presented in Fig. 2 (corresponds to Fig. 1 in the main text). We should note that the approximate results for Fock, thermal, and coherent states can be found in [2]. An analytic expression for the squeezed vacuum state Eq. (17), squeezed coherent state, and the correction to coherent state Eq. (15) are novel, to the best of our knowledge.

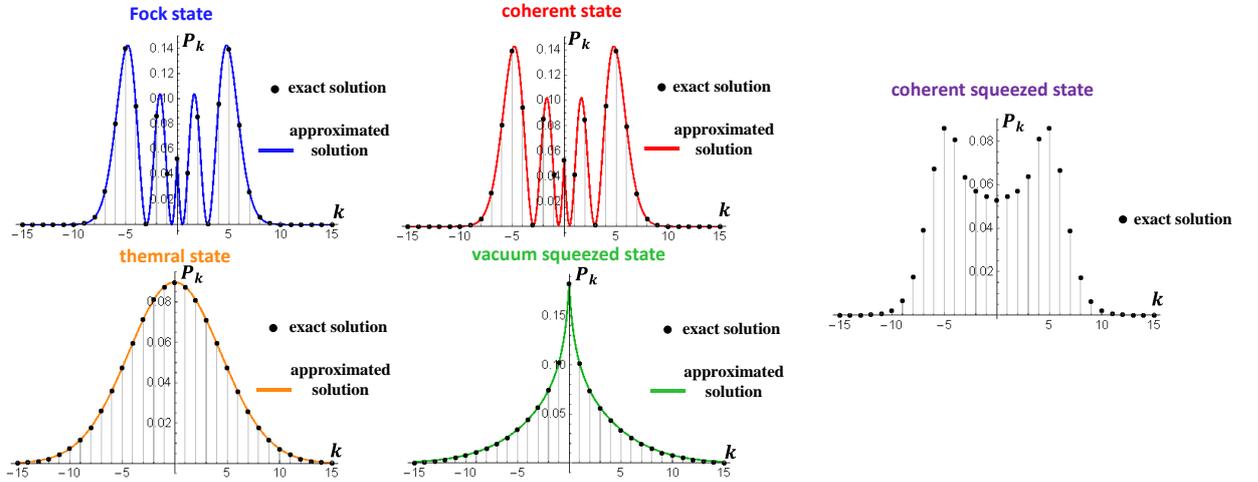

**Fig 2. The electron energy spectra after interaction with different photonic states. Comparing the approximate vs. the exact solutions**. The electron spectrum is calculated for coherent, Fock, thermal, squeezed vacuum, and squeezed coherent states. The black dots in each panel denote the exact solution calculated via Eq. (12). The blue, red, orange and green curves represent Eq. (14-17), respectively. For the squeezed coherent state, an analytical approximation was not found. All the plots are calculated for $\langle n \rangle = 1000$ and $|g| = 0.1$. For the coherent squeezed light, the parameter of squeezing is $|z| = 0.99$ and the phase $\phi = \pi/2$, i.e., phase-squeezed state.

**Section V – The fundamental connection between photon statistics and the electron energy spectrum**

This section is the key derivation of this work. Here we show a fundamental connection between the light's photon statistics and the free electron spectrum after the interaction with this light. We derive the exact equation which allows us to reconstruct the photon statistics from the probability peaks $P_k$ obtained using EELS. We start with the approximate Eq. (13) for the probabilities $P_k$:

$$P_k = \sum_{n=0}^{\infty} |J_k(2|g|\sqrt{n})|^2 p_n.$$

Let us consider only positive $k$, and then the Bessel function can be written in the following form:

$$J_k^2(2|g|\sqrt{n}) = \sum_{m=k}^{\infty} n^m c_{km},$$

where $c_{km} = \frac{(-1)^{m-k}(|g|^2)^m}{(m-k)!(m+k)!} \frac{(2m)!}{(m!)^2}$. Finally, we get the following expression for the spectrum:

$$P_k = \sum_{m=k}^{\infty} c_{km} \langle n^m \rangle, \qquad (18)$$

where $c_{km}$ is defined above, and $\langle n^m \rangle = \sum_{n=0}^{\infty} p_n n^m$ is the $m^{\text{th}}$ photon number moment. Eq. (18) corresponds to Eq. (4) of the main text. Eq. (18) is invertible, and hence we can find all the moments if we know all the $P_k$s:

$$\langle n^m \rangle = \sum_{k=m}^{\infty} d_{mk} P_k, \qquad (19)$$

where $d_{mk}$ are the required coefficients. Eq. (19) corresponds to Eq. (5) in the main texts. To find $d_{mk}$ we substitute Eq. (18) in Eq. (19):

$$\langle n^m \rangle = \sum_{k=m}^{\infty} d_{mk} P_k = \sum_{k=m}^{\infty} d_{mk} \left( \sum_{m=k}^{\infty} c_{kl} \langle n^l \rangle \right),$$

which leads to:

$$\sum_{k=m}^{\infty} d_{mk} \sum_{m=k}^{\infty} c_{kl} = \delta_{ml}, \qquad (20)$$

where $\delta_{ml} = \begin{cases} 1, m = l \\ 0, m \neq l \end{cases}$. Eq. (20) has an analytical solution for $d_{mk}$, which has the following form:

$$d_{mk} = 2^{\left\lfloor \frac{k-m+1}{2} \right\rfloor} \cdot \frac{k \cdot m!}{|g_{\text{q}}|^{2m}} \cdot \frac{\left(m - 1 + \left\lfloor \frac{k-m+1}{2} \right\rfloor\right)! \left(2m + 2\left\lfloor \frac{k-m}{2} \right\rfloor - 1\right)!!}{(k-m)!\,(2m-1)!!}, \qquad (21)$$

where $\lfloor ... \rfloor$ is the rounded-down integer part of the argument (floor function).

Eq. (19) and Eq. (21) give the ability to find all the photon number moments $\langle n^m \rangle$ if we know all $P_k$. If we know all the $\langle n^m \rangle$, then we can find the photon statistics $p_n$ in the majority of the cases [5]. Thus, here we derived the theory of free-electron quantum-optical detection (FEQOD). The ability to measure other photonic state properties and the accuracy of such precision are discussed in the next chapters. We should also note that all the results can be generalized to the exact Eq. (13), but the calculations are more involved and can only be done numerically.

**Section VI – Quantifying the precision of free-electron quantum-optical detection (FEQOD)**

In this section, we quantify the precision of FEQOD. Firstly, we calculate how the number of measurements affects the precision of the reconstructed moments $\langle n^m \rangle$. Finally, we analyze how the imperfections in the system resulting in the fluctuation of the coupling $g$ change the answer for $\langle n^m \rangle$. All these calculations justify the stability of the proposed measurement scheme.

To estimate the number of measurements we need to make in order to get a high enough precision, we analyze the following numerical model. We introduce a random variable, which can take discrete integer values $k$. The probability distribution of this variable being found in a specific $k$ is chosen to be $P_k$ of QPINEM theory. Using this numerical model, we obtain the spectrum with $N$ measurements and reconstruct the moments $\langle n^m \rangle$ from this spectrum. We find the relative deviation of the calculated moments from the exact solution. We average this relative deviation for different realizations and plot it in Fig. 3. According to Fig. 3, $N = 1000$ measurements give a relative deviation of 5%.

**(a) Different realizations of electron spectra using $N = 100$ measurements**

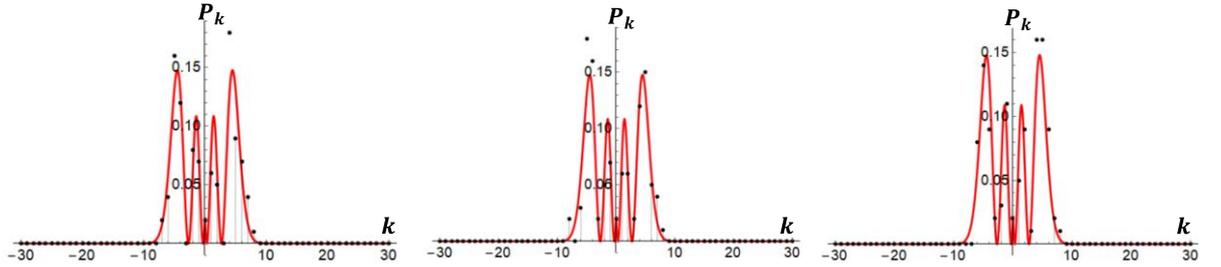

**(b) Average relative error in the first three moments after $N$ measurments**

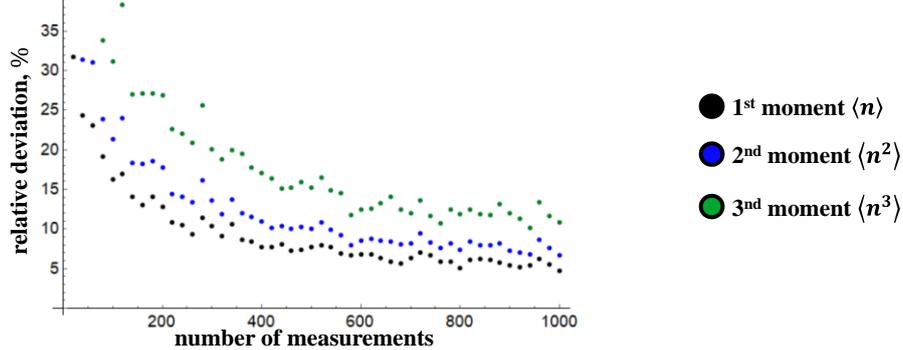

- 1st moment $\langle n \rangle$
- 2nd moment $\langle n^2 \rangle$
- 3nd moment $\langle n^3 \rangle$

**Fig 3. The precision of the FEQOD.** We perform calculations with $g = 0.1$ and coherent state $|\alpha\rangle$ with $\alpha = 30$. **(a)** We calculate several times the electron spectrum (using a random variable which has the same probability distribution as real PINEM) with $N$ measurements (black dots) and how it deviates from the exact solution (red curve). Here we plot three different realizations taken from the sample. **(b)** The spectrum is measured in 100 different realizations, using $N$ measurements. The averaged (in different realizations) relative deviation of the first three moments, reconstructed using Eq. (19), from the exact solution is calculated. The dependence of the relative deviation on the number of measurements is plotted for the 1st (black dots), 2nd (blue dots), 3rd (green dots) moments.

Now, let us estimate how the measurements depend on fluctuations in the coupling $|g|$. Fluctuation in the light's power, in the sample's position, in the electron's speed, and all other sources of fluctuation effectively result in the fluctuation of $|g|$. Hence, these calculations show the stability of the system in the general case. Numerical calculations are shown in Fig. 4

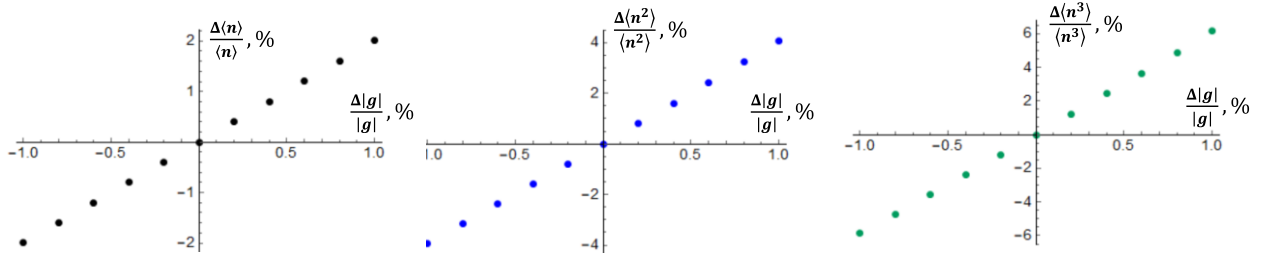

**Fig. 4 How a fluctuation in $|g|$ affects the measurements**. We evaluate the relative change in $\langle n^m \rangle$, plotted as $\frac{\Delta \langle n^m \rangle}{\langle n^m \rangle}$, as a function of the relative change in $|g|$, plotted as $\frac{\Delta |g|}{|g|}$. The calculation is done according to Eq. (19). The fluctuations are linear in $\Delta |g|$.

From Fig. 4, we see the following connection between an error in $|g|$ and in $\langle n^m \rangle$:

$$\frac{\Delta \langle n^m \rangle}{\langle n^m \rangle} \approx 2m \frac{\Delta |g|}{|g|}, \qquad (22)$$

where $\Delta |g|$ is the deviation of interaction strength from its average value $|g|$; $\Delta \langle n^m \rangle$ is the deviation from the average value of $\langle n^m \rangle$. This calculation shows how the measurements depend on the fluctuation in the parameters of the system.

**Section VII – Quantifying the non-destructive nature of the measurement: tracing out the electron.**

This section considers the change to the measured photonic state when perturbed by the interaction with an electron. We quantify the non-destructive nature of the measurement by tracing-out or post-selecting the joint density matrix, and checking the change to the resulting photonic density matrix. Here, we show that in the regime of parameters when Eq. (11) holds, the electron interaction *does not* change the photon number distribution. In a later section, we also investigate the sensitivity of squeezed light to any decoherence resulting from the light-electron interaction and show that it also experiences only a minute change. This analysis, therefore, demonstrates the non-destructive nature of the measurement.

We consider the following scheme: a single free electron interacts with the light, and the electron energy is then measured. We carry out these measurements for an ensemble of $N$ identical photonic states and measure how, on average (i.e., after tracing out all possible outcomes for the electron energy), the photon statistics changes after one interaction. The answer to this problem can be found using the joint probabilities $P_{nk}$ from Section II:

$$p'_n = \sum_{k=-\infty}^{\infty} P_{nk}, \qquad (23)$$

where $P_{nk}$ are the probabilities of finding $n$ photons and an electron with energy $E_0 + k\hbar\omega$. In the range of parameters where $P_{nk}$ can be described by the approximated Eq. (11), we have:

$$p'_n = \sum_{k=-\infty}^{\infty} P_{nk} = \sum_{k=-\infty}^{+\infty} |J_k(2|g|\sqrt{n})|^2 p_n = p_n, \qquad (24)$$

where $p_n$ is the photon statistics before the interaction and $p'_n$ is the photon statistics after the interaction. Here we use the following identity: $\sum_{k=-\infty}^{+\infty} |J_k(x)|^2 = 1$.

Eq. (11) holds for most experiments; however, let us analyze the change in photon number distribution for the exact Eq. (10) (strong coupling). In this case, the photonic state is modified, and we quantify this change for the coherent state $|\alpha\rangle$ in Fig. 5. The coherent state is just an example that illustrates the central concept, but indeed similar calculations can be done for the most general photonic state. Fig. 10 shows how the photon statistics are modified and how this modification depends on the coupling $g$.

To better quantify the change in the photon statistics, we calculate the maximum deviation in the distribution after one interaction $\max_n |p'_n - p_n|$ as a function of $|g|$ for a different number of photons (see Fig. 5).

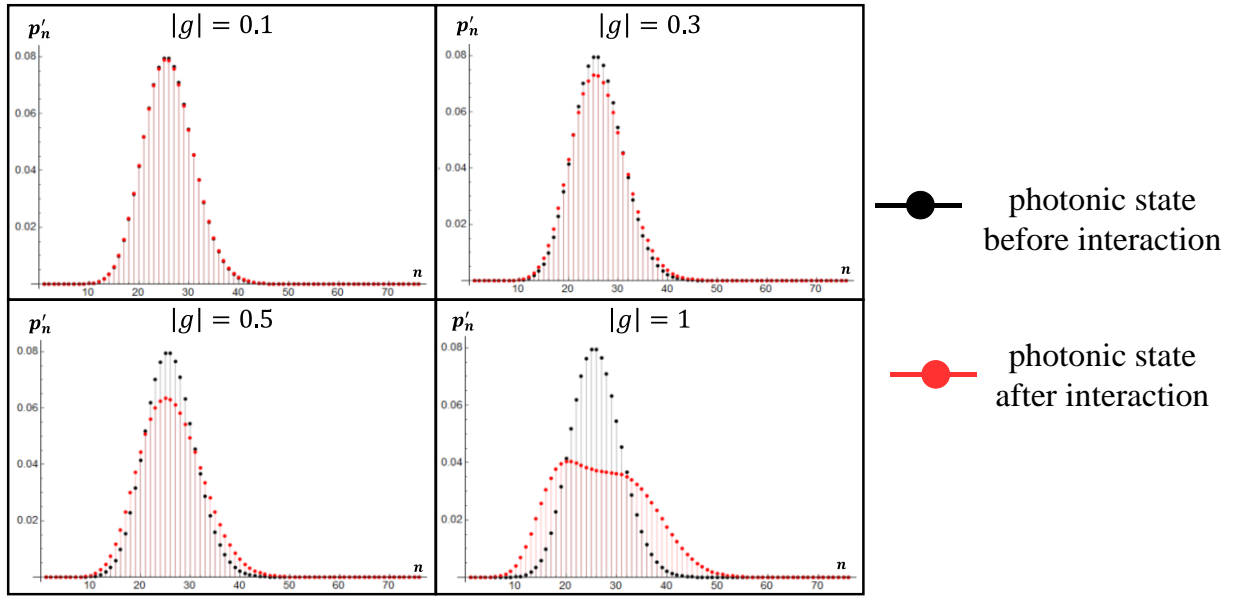

**Fig 5. The photon statistics change due to the interaction with a free electron** (after tracing-out over the electron). We plot the photon statistics of the coherent state $|\alpha\rangle$ (with $\alpha = 5$) before (black dots) and after (red dots) the interaction with a free electron for different coupling strengths $|g|$. We can see that if $|g| < 0.3$, the change in the statistics is small. The calculation is made using the exact equation (Eq. 10).

Fig. 5 justifies the analytic Eq. (24). We can see that if we increase the number of photons and decrease $|g|$ (it is precisely the conditions when Eq. (24) is valid), we can make the change in the photon statistics as small as we want.

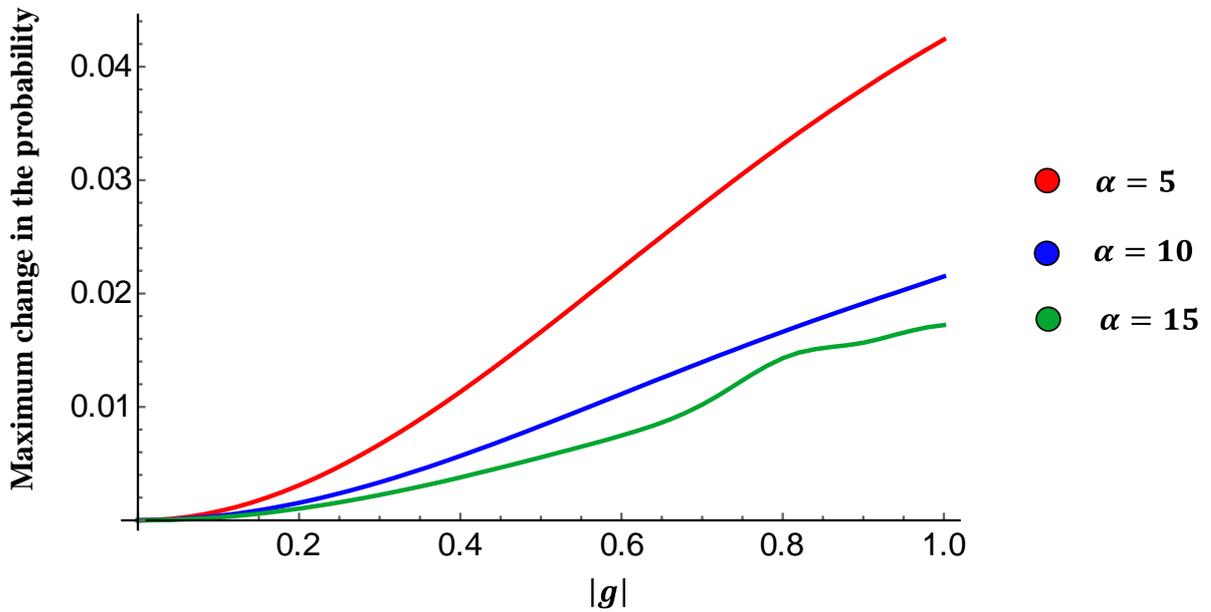

**Fig 6. The change in photon statistics due to the interaction with a free electron as a function of $|g|$** (after tracing-out over the electron). We plot the maximum change of photon statistics $\max_n |p'_n - p_n|$ of the coherent state $|\alpha\rangle$ as a function of $|g|$ for different $\alpha = 5, 10, 20$, red, blue, and green curves, respectively. The calculation is done using the exact equation (Eq. 10).

**Section VIII – Quantifying the non-destructive nature of the measurement: postselected electron's energy**

In this section, we calculate how the photonic state changes if we measure the energy of the electron after the interaction and it equals to $E_k$. We calculate the photon statistics after the interaction:

$$p'_{nk} = \frac{P_{nk}}{\sum_n P_{nk}} = \frac{P_{nk}}{P_k}, \qquad (25)$$

where $P_{nk}$ and $P_k$ should be calculated according to Eq. (10) and Eq. (12). Fig. 7, calculated for the coherent state, shows how the change in the statistics depends on the measured energy $E_k$.

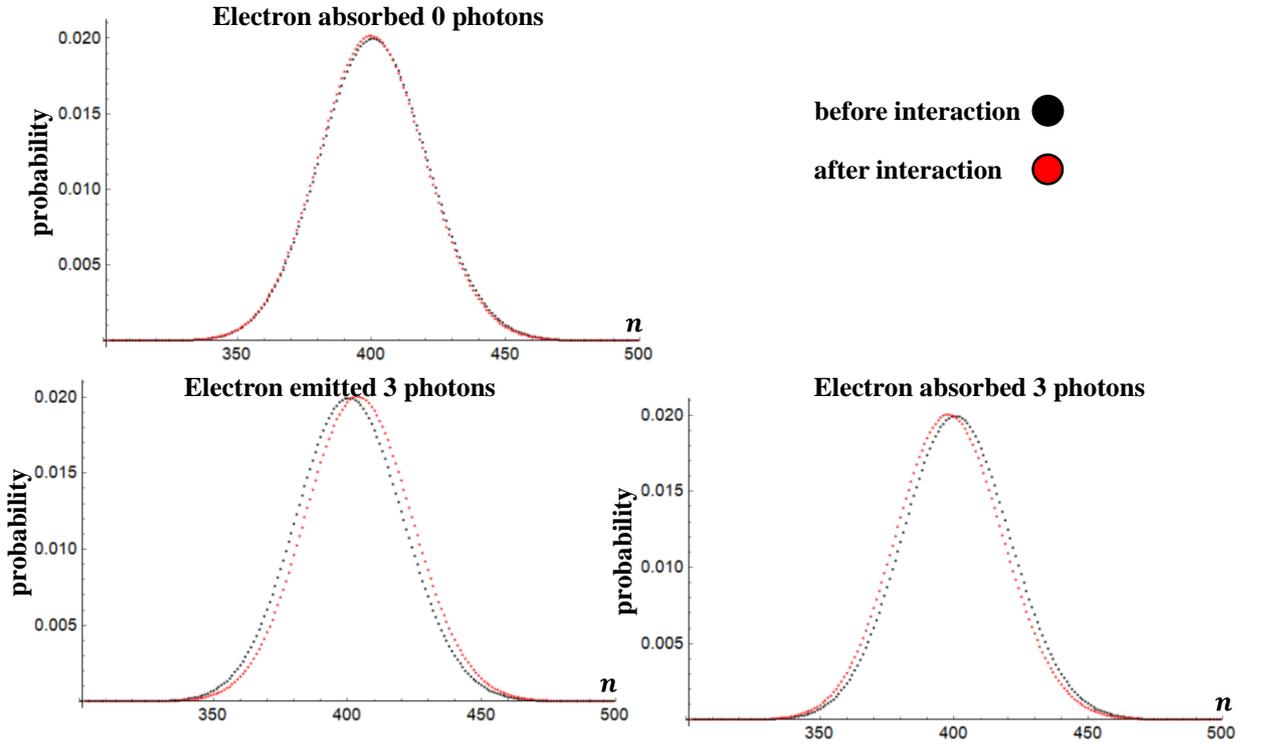

**Fig 7. The change in photon statistics due to the interaction with a free electron as a function of $|g|$** (after measurement of the electron). We plot the change in photon statistics of the coherent state $|\alpha\rangle$ with $\alpha = 20$ and coupling $|g| = 0.1$. We consider three different situations: the measured electron absorbed 0 photons, three photons, and emitted three photons. The calculation is done using Eq. 25.

As the result of these two sections, we can conclude that if the interaction constant is not too large ($g < 0.1$), then a single electron will not change the photon statistics significantly in the case when it is traced out as well as in the case when it is measured.

**Section IX – How multiple electrons that interact with a single light pulse modify the state of light: testing the feasibility of the single-shot measurement scheme.**

In this section, we estimate the change of the photonic state after the interaction with $N_e$ independent electrons. We make the following model: electrons interact with the photonic state sequentially, one after the other. Each electron modifies the photonic state and is then traced out. We calculate the maximum change in the statistics as in the previous section after the interaction with $N_e$ electrons (Fig. 8).

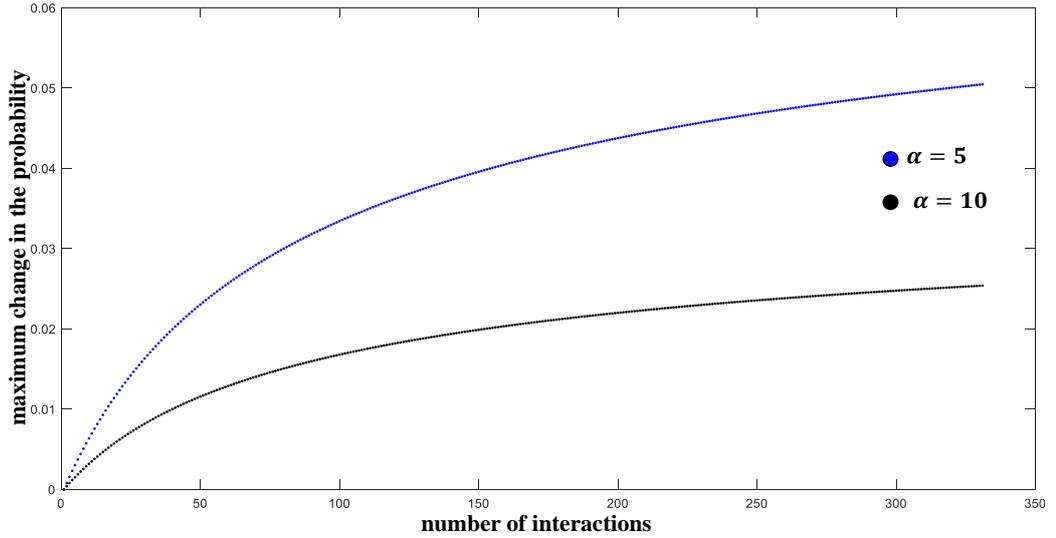

Fig 8. **The maximal change in photon statistics after the interaction with $N_e$ electrons**. We calculated the maximal change of photon statistics $\max_n |p'_n - p_n|$ of the coherent state $|\alpha\rangle$ with $\alpha = 5$ (blue dots) and $\alpha = 10$ (black dots) depending on the number of electrons $N_e$ interacted with this light. The change in the statistics has slower than linear growth with $N_e$.

According to Fig. 8, the change in the photon statistics grows with the increase of $N_e$. However, the rate of this growth is slower than linear. Moreover, the change in the photon statistics decreases with the increase in the number of photons. Hence, we can conclude that if the coupling strength $g$ is small and the number of photons is large enough, then the photon statistics of the light interacting with an $N_e$-electron beam does not change significantly. This conclusion justifies the *single-shot* measurement (measurement with a single many-electron pulse) for intense photonic states and weak coupling.

**Section X – The effect of electron interactions on the off-diagonal elements of the photonic density matrix. The case of squeezed state.**

In Sections VII, VIII, and IX, we estimated how the electron interaction changes the photon statistics, namely, the diagonal density matrix elements. However, some important observables in quantum optics emerge from the off-diagonal elements and are thus prone to decoherence. Thus, the claim of non-destructive measurements of the entire photonic state has to be complemented with the analysis of such observables as well. For example, we consider the case of a squeezed state and estimate the change in the squeezing after interaction with $N$ consequent electrons.

We calculate the variance of the amplitude quadrature before and after the interaction:

$$\Delta X^2 = \langle X^2 \rangle - \langle X \rangle^2,$$

where $X = \frac{1}{2}(a + a^\dagger)$. For a squeezed state [8]: $\Delta X^2 = \frac{1}{4}e^{-2r}$, where $r$ is the squeeze parameter. After interaction with one electron:

$$\langle X \rangle' = \text{Tr}[\rho_{\text{ph}} \cdot S^\dagger X S] = \frac{1}{2}\text{Tr}[\rho_{\text{ph}} \otimes \rho_e \cdot (a + gb + a^\dagger + g^* b^\dagger)] = \langle X \rangle,$$

here we used $S^\dagger a S = a + gb$ and $\text{Tr}[\rho_e b] = 0$, which holds for the monoenergetic electron. Then, the variance in quadrature after the interactions can be written:

$$(\Delta X^2)' = \langle X^2 \rangle' - (\langle X \rangle')^2 = \frac{1}{4}\text{Tr}[\rho_{\text{ph}} \otimes \rho_e \cdot S^\dagger (a + a^\dagger)^2 S] - \langle X \rangle^2$$

$$= \frac{1}{4}\text{Tr}[\rho_{\text{ph}} \otimes \rho_e \cdot (a + gb + a^\dagger + g^* b^\dagger)^2 S] - \langle X \rangle^2 = \Delta X^2 + \frac{1}{2}|g|^2$$

Hence, after one interaction, the variance of the $X$ quadrature equals:

$$(\Delta X^2)' = \Delta X^2 + \frac{1}{2}|g|^2, \tag{26}$$

where $(\Delta X^2)'$ is the quadrature after interaction and $\Delta X^2$ is the quadrature before interaction. Hence, according to Eq. (26), for non-destructive measurements with $|g| \ll 1$, the quality of squeezing changes negligibly, and so do the off-diagonal elements of the photonic density matrix. Thus, this example complements our previous analysis and shows that the non-destructive QPINEM measurement is expected to change the entire density matrix of the photonic state negligibly.

Let us consider a single-shot measurement, where the photonic state interacts with $N_e$ consequent electrons. In this case, the change in the quadrature variance equals to:

$$(\Delta X^2)' = \Delta X^2 + \frac{1}{2} N_e |g|^2. \tag{27}$$

We need a large enough number of electrons $N_e$ to measure the state with small enough relative error (estimated in Fig. 3b). On the other hand, if we increase the number of electrons $N_e$ in a single-shot measurement, we will modify the photonic state significantly. Thus, the single-shot measurement provides only a destructive measurement (i.e., significantly modifies the photonic states), otherwise resulting in a large relative error: a tradeoff signature between measurement accuracy and perturbation of the observable being measured. However, if one tolerates a destructive measurement, the entire measurement time is extremely small and enables measurement of the statistics in a single-shot, i.e., without repeating the experiment.

**Section XI – Quantum tomography of light based on free electrons**

The measurement scheme discussed in previous sections gives the possibility to measure photon statistics. However, in many cases, it is crucial to find the photonic state's entire density matrix. In Fig 9, the schemes for measuring photon statistics and state reconstruction (i.e., quantum tomography) are depicted. The quantum tomography scheme (Fig 9 (b)) is based on a two-point interaction. In the first point, a free electron is modified by the intense classical light of the laser, i.e., the local oscillator (LO). In the second point of interaction, this same electron interacts with the quantum light we want to measure. We measure the dependence of the electron's spectrum $P_k(\theta)$ on the phase of the LO $\theta$ (Fig. 4 in the main text). Using Eq. (19), it is possible to reconstruct the photon statistics $p_n(\theta)$ for the fixed $\theta$. By changing $\theta$, we can completely reconstruct the photonic state using the Radon transformation [9].

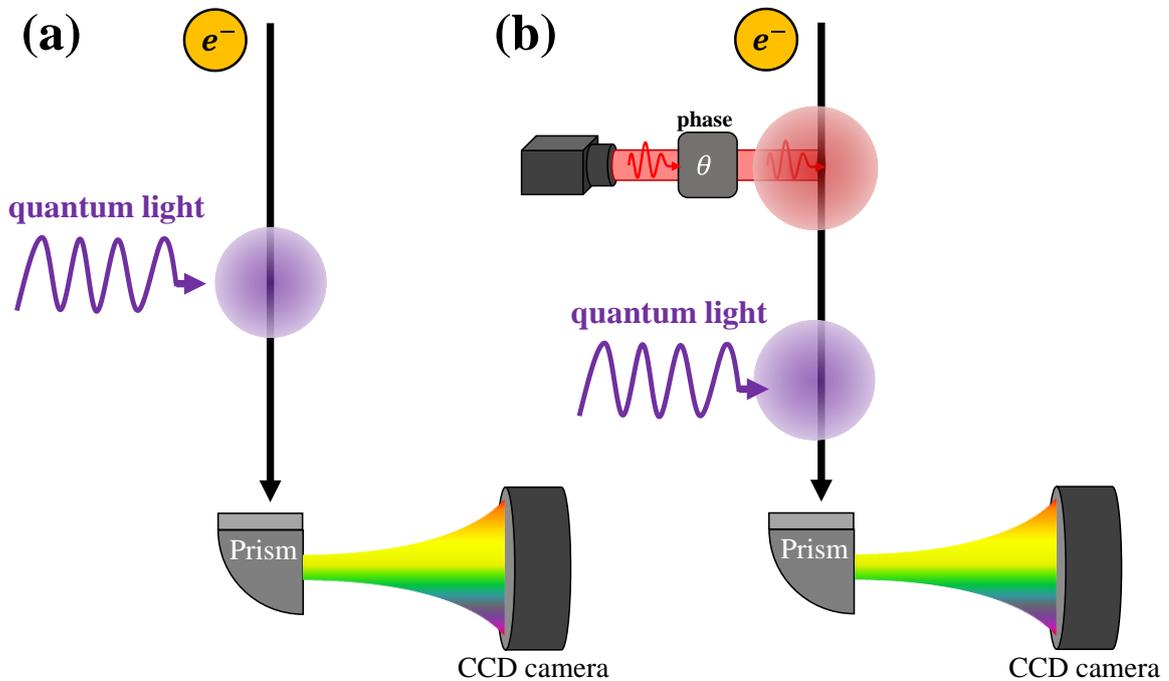

**Fig 9. Free-electron-based quantum tomography of photonic states**. (**a**) The basic scheme considered in the previous sections is where we have one point of interaction and measure the energy spectrum of electrons after the interaction. In this way, we can extract the photon statistics of the quantum light but cannot extract other photonic properties. (**b**) A more advanced scheme that we suggest in this section. This scheme enables quantum tomography based on free electrons. There are two points of interaction, the first one is with a local oscillator (laser) with phase $\theta$ which can be changed, and the second one is with the quantum light that we want to reconstruct.

The concept of FEQOD is very general. As an example, let us consider the field quadrature, which is defined by the equation:

$$X(\theta) = \frac{1}{2}(ae^{-i\theta} + a^\dagger e^{i\theta}).$$

According to Section I, the photon annihilation operator after two consecutive interactions is defined via:

$$\tilde{a} = \frac{1}{\sqrt{2}}(a + a_{LO}) \cong \frac{1}{\sqrt{2}}(a + \alpha)$$

where we replaced the LO operator with its c-number amplitude $\alpha = |\alpha|e^{-i\theta}$ on account of it being a strong classical state, giving:

$$n = \frac{1}{2}(|\alpha|e^{-i\theta} + a^\dagger)(|\alpha|e^{i\theta} + a),$$

where $\alpha$ is the parameter of the coherent state of the laser (LO), and $\theta$ is its phase, $a$ is annihilation operator for the quantum state that we want to measure. It can be simplified according to:

$$n = \frac{1}{2}\left(|\alpha|^2 + a^\dagger a + 2|\alpha|X(\theta)\right). \tag{28}$$

According to Eq. (19), we can find the moments $\langle n^m \rangle$ and then, Eq. (28) defines $n$ in terms of the quadrature $X(\theta)$, which gives the possibility to measure the distribution of $X(\theta)$. These distributions are of great importance and are used for reconstructing the Wigner function of the measured light, for example, when measuring the state's squeezing [6, 7].

**Section XII – Measurement of the temporal degrees of coherence of quantum light**

This section considers the case when the light is almost monochromatic but still has some spectral width. In this case, we can investigate the temporal coherence of such light. Here we again employ a two-point interaction scheme similar to Fig. 9(b) and show how we can reconstruct the temporal degree of quantum coherence from the electron's spectrum.

Let us start with a two-point interaction scheme. The quantum light is split into two parts, each with an equal number of photons (Fig 10). One part of the light is delated by time $\tau$ compared with the other. Free electrons interact with both parts of light, and we measure the resulting spectrum.

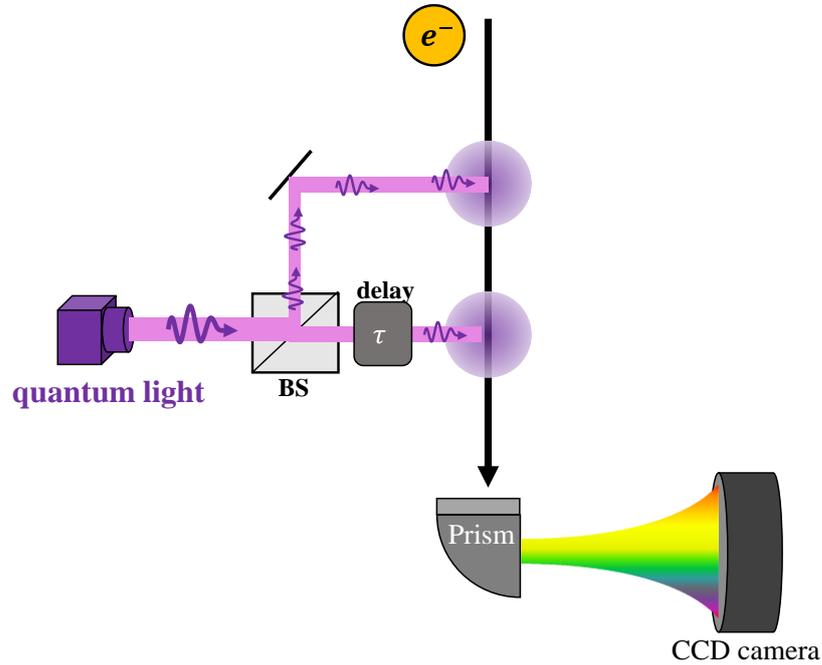

**Fig 10. Free-electron quantum-optical detection of the temporal coherence of light**. The source of light for which we want to measure the temporal coherence is split into two parts. One part is delayed by time $\tau$ relative to the other part. There could be a distance between the two points of interaction, which could be considered in the analysis. The electron interacts with both parts of the light, and the resulting electron energy spectrum is measured.

Usually, the second-order temporal coherence of the light is defined by the introduction of the normalized second-order coherence function $g^{(2)}$ [6, 7, 8]:

$$g^{(2)}(\tau) = \frac{\langle a^\dagger(t) a^\dagger(t+\tau) a(t+\tau) a(t)\rangle}{\langle a^\dagger(t) a(t)\rangle^2}, \qquad (29)$$

where $a(t)$ and $a^\dagger(t)$ the annihilation and creation operators for light having a finite bandwidth, which can be found in [8]. We redefine annihilation operator as $a'(t) = \frac{1}{2}(a(t) + a(t+\tau))$ and then the photon number operator becomes:

$$n(\tau) = \frac{1}{4}\left(a^\dagger(t+\tau) + a^\dagger(t)\right)\left(a(t+\tau) + a(t)\right). \tag{30}$$

Please, note that the definition here differs from Section I in $\sqrt{2}$ times, because here the operators $a(t)$ and $a(t+\tau)$ are dependent. We can find the electron spectrum $P_k(\tau)$, which depends on the delay (Fig. 5 in the main text), and then according to Eq. (19), we can find all the correlation moments $\langle n^m(\tau)\rangle$, where the number operator is defined in Eq. (30). Then, we can introduce a new observable having similar properties to the conventional second-order coherence $g^{(2)}$, and also quantifies the second-order temporal coherence:

$$\tilde{g}^{(2)}(\tau) = 2\frac{\langle n^2(\tau)\rangle - \langle n(\tau)\rangle}{\langle n(\tau)\rangle^2} - \frac{\langle n^2(0)\rangle - \langle n(0)\rangle}{\langle n(0)\rangle^2}, \tag{31}$$

where $\tilde{g}^{(2)}(\tau)$ is the new measure of coherence (used in Fig. 5 of the main text), and $n(\tau)$ is defined according to Eq. (30). Let us investigate the properties of $\tilde{g}^{(2)}(\tau)$. It can be shown that:

$$\tilde{g}^{(2)}(0) = g^{(2)}(0). \tag{32}$$

Moreover, for the long delay $\tau \to \infty$, when the coherence disappears, it can be shown that:

$$\tilde{g}^{(2)}(\tau \to \infty) = 2\left(1 - \frac{1}{\langle n\rangle}\right). \tag{33}$$

Eq. (33) shows the main difference between $\tilde{g}^{(2)}$ and $g^{(2)}$ because $g^{(2)}(\tau \to \infty) = 1$. It means that the plot for $\tilde{g}^{(2)}(\tau)$ for coherent and thermal states will have the form as sketched in Fig. 8. However, the temporal width of $g^{(2)}$ and $\tilde{g}^{(2)}$ is the same.

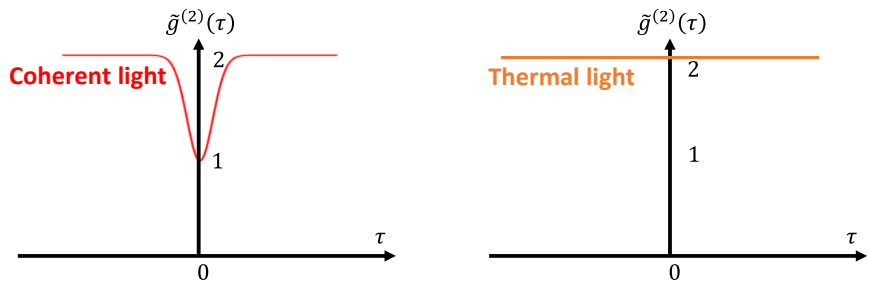

**Fig 8. The predicted temporal coherence $\tilde{g}^{(2)}$ for coherent and for thermal light.** The $\tilde{g}^{(2)}$ is defined in Eq. (29); it has the opposite behavior to $g^{(2)}$ (defined in Eq. (27)). While $\tilde{g}^{(2)}$ is almost constant for thermal light; it has a dip for coherent light. Furthermore, $\tilde{g}^{(2)}$ tends to 2 for $\tau \to \infty$ ($g^{(2)}$ has the opposite behavior and tends to 1 for $\tau \to \infty$).

**References**

[1] O. Kfir, Entanglements of Electrons and Cavity Photons in the Strong-Coupling Regime, *Phys. Rev. Lett.* **123**, 103602 (2019).

[2] V. Giulio, M. Kociak, & F. J. G. de Abajo, Probing quantum optical excitations with fast electrons. *Optica* **6**, 1524–1534 (2019).

[3] R. Ruimy, A. Gorlach, C. Mechel, N. Rivera, I. Kaminer, Towards atomic-resolution quantum measurements with coherently-shaped electrons, arXiv:2011.00348

[4] Z. Zhao, X.-Q. Sun, S. Fan, Quantum entanglement and modulation enhancement of free-electron−bound-electron interaction, arXiv:2010.11396

[5] V. John, I. Angelov, A. A. Oncul, D. Thevenin, Techniques for the representation of a distribution from a finite number of its moments, *Chem. Eng. Sci.* **62** (11), 2890-2904 (2007).

[6] M. O. Scully, M. S. Zubairy, *"Quantum Optics"* (Cambridge University Press, New York, 1997).

[7] L. Mandel, E. Wolf, *Optical Coherence and Quantum Optics*, (Cambridge University Press, New York 1995).

[8] C. Gerry and P. Knight, *Introductory Quantum Optics*, (Cambridge University Press, New York 2004).

[9] U. Leonhardt, *Measuring the Quantum State of Light*, (Cambridge University Press, New York 1997).